\documentclass[10pt,twocolumn,letterpaper]{article}

\usepackage{cvpr}
\usepackage{times}
\usepackage{epsfig}
\usepackage{graphicx}
\usepackage{amsmath}
\usepackage{mathtools}
\usepackage{threeparttable}
\usepackage{amssymb}
\usepackage{multirow}
\usepackage{algorithm}
\usepackage{balance}
\usepackage[noend]{algpseudocode}
\usepackage{subcaption}
\makeatletter
\def\algbackskip{\hskip-\ALG@thistlm}
\makeatother

\usepackage[pagebackref=true,breaklinks=true,letterpaper=true,colorlinks,bookmarks=false]{hyperref}

\cvprfinalcopy 

\def\cvprPaperID{6435} 

\ifcvprfinal\fi

\begin{document}

\title{Enhancing Cross-task Black-Box Transferability of Adversarial Examples\\ with Dispersion Reduction}

\author{
Yantao Lu\thanks{Equal contribution}$\ ^1$, Yunhan Jia\textsuperscript{*}$^2$, Jianyu Wang$^3$, Bai Li$^4$, Weiheng Chai$^1$, Lawrence Carin$^4$, Senem Velipasalar$^1$\\
$^1$Syracuse University, $^2$University of Michigan, $^3$UCLA, $^4$Duke University \\
\{ylu25, wchai01, svelipas\}@syr.edu, \{jack0082010, wjyouch\}@gmail.com, \{bai.li, lcarin\}@duke.edu \\
} 

\maketitle

\begin{abstract}
Neural networks are known to be vulnerable to carefully crafted adversarial examples, and these malicious samples often transfer, i.e., they remain adversarial even against other models. Although great efforts have been delved into the transferability across models, surprisingly, less attention has been paid to the cross-task transferability, which represents the real-world cybercriminal's situation, where an ensemble of different defense/detection mechanisms need to be evaded all at once. In this paper, we investigate the transferability of adversarial examples across a wide range of real-world computer vision tasks, including image classification, object detection, semantic segmentation, explicit content detection, and text detection. Our proposed attack minimizes the ``dispersion'' of the internal feature map, which overcomes existing attacks' limitation of requiring task-specific loss functions and/or probing a target model. We conduct evaluation on open source detection and segmentation models as well as four different computer vision tasks provided by Google Cloud Vision (GCV) APIs, to show how our approach outperforms existing attacks by degrading performance of multiple CV tasks by a large margin with only modest perturbations ($l_\infty\leq16$).

\end{abstract}
\section{Introduction}
Recent progress in adversarial machine learning has brought the weaknesses of deep neural networks (DNNs) into the spotlight, and drawn the attention of researchers working on security and machine learning. Given a deep learning model, it is easy to generate adversarial examples (AEs), which are close to the original input, but are easily misclassified by the model~\cite{carlini2017towards, szegedy2013intriguing}. More importantly, their effectiveness sometimes \emph{transfers}, which may severely hinder DNN-based applications especially in security critical scenarios~\cite{liu2016delving, dong2018boosting, xie2018improving}. While such problems are alarming, little attention has been paid to the threat model of commercially deployed vision-based systems, wherein deep learning models across different tasks are assembled to provide fail-safe protection against evasion attacks. Such a threat model turns out to be quite different from those models that have been intensively studied by aforementioned research.

\begin{figure}
\centering
\includegraphics[width=\columnwidth]{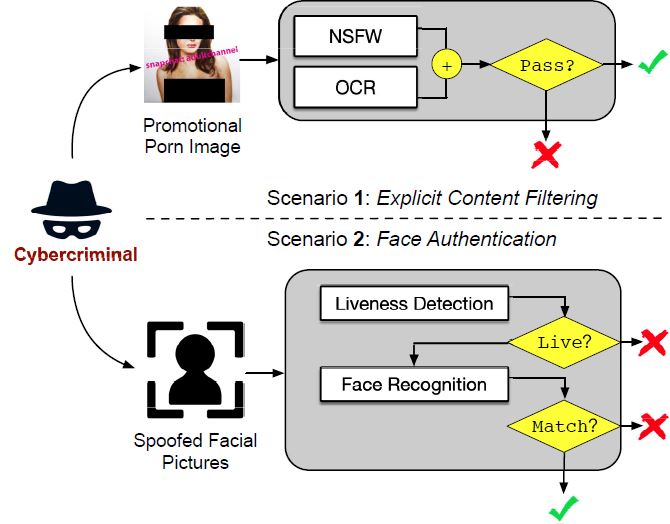}
\vspace{-0.5cm}
\caption{
\small
Real-world computer vision systems deployed in safety- and security-critical scenarios usually employ an ensemble of detection mechanisms that are opaque to attackers. Cybercriminals are required to generate adversarial examples that transfer across tasks to maximize their chances of evading the entire detection systems.}
\label{fig:threat}
\end{figure}

\textbf{Cross-task threat model.} Computer vision (CV) based detection mechanisms have been deployed extensively in security-critical applications, such as content censorship and authentication with facial biometrics, and readily available services are provided by cloud giants through APIs (e.g., Google Cloud Vision~\cite{google}). The detection systems have long been targeted by evasive attacks from cybercriminals, and it has resulted in an arm race between new attacks and more advanced defenses.~To overcome the weakness of deep learning in an individual domain, real-world CV systems tend to employ an ensemble of different detection mechanisms to prevent evasions. As shown in Fig.~\ref{fig:threat}, underground businesses embed promotional contents such as URLs into porn images with sexual content for illicit online advertising or phishing. A detection system, combining Optical Character Recognition (OCR) and image-based explicit content detection, can thus drop posted images containing either suspicious URLs or sexual content to mitigate evasion attacks. Similarly, a face recognition model that is known to be fragile~\cite{sharif2016accessorize} is usually protected by a liveness detector to defeat spoofed digital images when deployed for authentication. Such ensemble mechanisms are widely adopted in real-world CV deployment.

To evade detection systems with uncertain underlying mechanisms, attackers turn to generating adversarial examples that transfer across CV tasks. Many adversarial techniques on enhancing transferability have been proposed~\cite{zhou2018transferable, xie2018improving, liu2016delving, dong2018boosting}. However, most of them are designed for image classification tasks, and rely on task-specific loss functions (e.g., cross-entropy loss), which limits their effectiveness when transferred to other CV tasks.

\begin{figure}
\vspace{-0.5cm}
\centering
\includegraphics[width=\columnwidth]{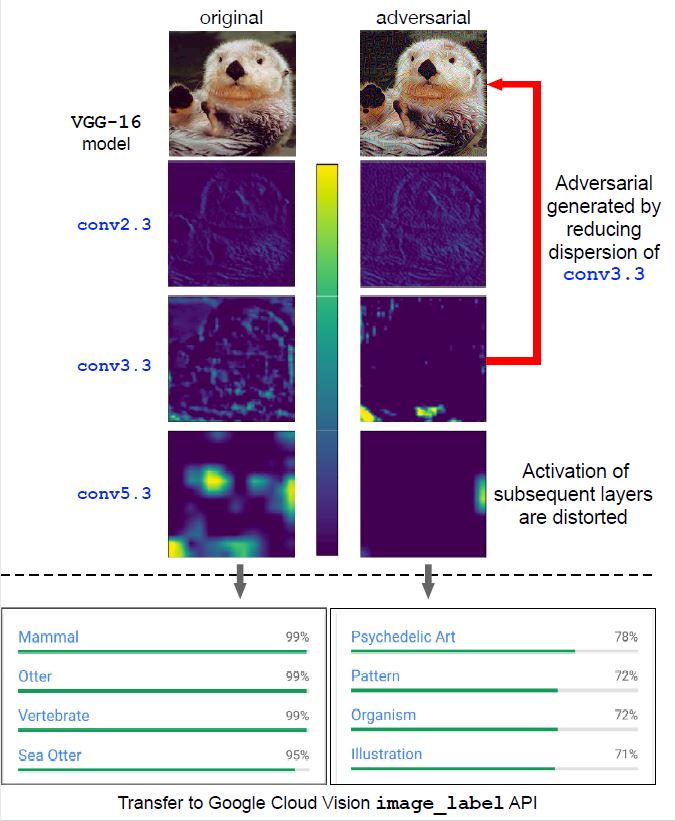}
\caption{
\small
DR attack targets on the dispersion of feature map at a specific layer of feature extractors. The adversarial example generated by minimizing dispersion at \texttt{conv3.3} of VGG-16 model also distorts feature space of subsequent layers (e.g., \texttt{conv5.3}), and its effectiveness transfers to commercially deployed GCV APIs.}
\label{fig:illustration}
\end{figure}

To provide a strong baseline attack to evaluate the robustness of DNN models under the aforementioned threat model, we propose a new succinct method to generate adversarial examples, which transfer across a broad class of CV tasks, including classification, object detection, semantic segmentation, explicit content detection, and text detection and recognition. Our approach, called \emph{Dispersion Reduction} (\textbf{DR}) and illustrated in Fig.~\ref{fig:illustration}, is inspired by the impact of ``contrast'' on an image's perceptibility. As lowering the contrast of an image would make the objects indistinguishable, we presume that reducing the ``contrast'' of an internal feature map would also degrade the recognizability of objects in the image, and thus could evade CV-based detections.

We use \emph{dispersion} as a measure of ``contrast'' in feature space, which describes how scattered the feature map of an internal layer is. We empirically validate the impact of dispersion on model predictions, and find that reducing the dispersion of internal feature map would largely affect the activation of subsequent layers. Based on additional observation that lower layers detect simple features~\cite{lee2009convolutional}, we hypothesize that the low level features extracted by early convolution layers share many similarities across CV models. By reducing the dispersion of an internal feature map, the information that is in the feature output becomes indistinguishable or useless, and thus the following layers are not able to obtain any useful information no matter what kind of CV task is at hand. Thus, the distortions caused by dispersion reduction in feature space, are ideally suited to fool any CV model, whether designed for classification, object detection, semantic segmentation, text detection, or other vision tasks.

Based on these observations, we propose and build the \textbf{DR} as a strong baseline attack to evaluate model robustness against black box attacks, which generate adversarial examples using simple and readily-available image classification models (e.g., VGG-16, Inception-V3 and ResNet-152), whose effects extend to a wide range of CV tasks. We evaluate our proposed DR attack on both popular open source detection and segmentation models, as well as commercially deployed detection models on four Google Cloud Vision APIs: classification, object detection, SafeSearch, and Text Detection (see \S\ref{sec:experiments}). ImageNet, PASCAL VOC2012 and MS COCO2017 datasets are used for evaluations. The results show that our proposed attack causes larger drops on the model performance compared to the state-of-the-art attacks ( MI-FGSM~\cite{dong2018boosting}, DIM~\cite{xie2018improving} and TI~\cite{dong2019evading}) across different tasks. We hope our finding to raise alarms for real-world CV deployment in security-critical applications, and our simple yet effective attack to be used as a benchmark to evaluate model robustness. Code is available at: \href{https://github.com/anonymous0120/dr}{https://github.com/anonymous0120/dr}.

\textbf{Contributions.} The contributions of this work include the following:
\vspace{-\topsep}
\begin{itemize}

\item This work is the first to study adversarial machine learning for cross-task attacks. The proposed attack, called dispersion reduction, does not rely on labeling systems or task-specific loss functions.
\vspace{-0.3cm}
\item Evaluation shows that the proposed DR attack beats state-of-the-art attacks in degrading the performance of object detection and semantic segmentation models and four different GCV API tasks by a large margin: 52\% lower mAP (detection) and 31\% lower mIoU (segmentation) compared to the best of the baseline attacks.
\vspace{-0.3cm}
\item Code and evaluation data are all available at an anonymized GitHub repository~\cite{our-code}.
\end{itemize}

\section{Related Work}
Adversarial examples~\cite{szegedy2013intriguing,goodfellow2014explaining} have recently been shown to be able to transfer across models trained on different datasets, having different architectures, or even designed for different tasks~\cite{liu2016delving,xie2017adversarial}. This transferability property motivates the research on black-box adversarial attacks.

One notable strategy, as demonstrated in~\cite{papernot2017practical,papernot2016transferability}, is to perform black-box attacks using a substitute model, which is trained to mimic the behavior of the target model by distillation technique. They also demonstrated black-box attacks against real-world machine learning services hosted by Amazon and Google. Another related line of research, a.k.a. gradient-free attack, uses feedback on query data, i.e. soft predictions~\cite{uesato2018adversarial,ilyas2018black} or hard labels~\cite{brendel2017decision} to construct adversarial examples.

The limitation of the aforementioned works is that they all require (some form of) feedback from the target model, which may not be practical in some scenarios. Recently, several methods have been proposed to improve transferability by studying the attack generation process itself, and our method falls into this category. In general, an iterative attack~\cite{carlini2017towards,kurakin2016adversarial,pgd} achieves higher attack success rate than a single-step attack~\cite{goodfellow2014explaining} in white-box setting, but performs worse when transferred to other models.~The methods mentioned below reduce the overfitting effect by either improving the optimization process or by exploiting data augmentation.



\textbf{MI-FGSM.} Momentum Iterative Fast Gradient Sign Method (MI-FGSM)~\cite{dong2018boosting} integrates momentum term into the attack process to stabilize update directions and escape poor local maxima. The update procedure is as follows:
\begin{equation}
\begin{aligned}
x'_{t+1} = x'_t + \alpha \cdot sign(g_{t+1}) \\
g_{t+1} = \mu \cdot g_{t} + \frac{\bigtriangledown_{x}J(x'_{t},y)}{ \parallel \bigtriangledown_{x}J(x'_{t},y) \parallel_1}
\end{aligned}
\label{eq:mifgsm}
\end{equation}
The strength of MI-FGSM can be controlled by the momentum and the number of iterations. 

\textbf{DIM.} Momentum Diverse Inputs Fast Gradient Sign Method (DIM) combines momentum and input diversity strategy to enhance transferability~\cite{xie2018improving}. Specifically, DIM applies image transformation, $T(\cdot)$, to the inputs with a probability $p$ at each iteration of iterative FGSM to alleviate the overfitting phenomenon. The update procedure is similar to MI-FGSM, the only difference being the replacement of  Eq.~(\ref{eq:mifgsm}) by:
\begin{equation}
\begin{aligned}
x'_{t+1} = Clip^\epsilon_x\{x'_t + \alpha \cdot sign(\bigtriangledown_x L(T(x'_{t+1};p),y^{true})\} \\
\end{aligned}
\label{eq:dim}
\end{equation}
where $T(x'_t,p)$ is a stochastic transformation function that performs input transformation with probability $p$.

\textbf{TI.} Rather than optimizing the objective function at a single point, Translation-Invariance (TI)~\cite{TI} method uses a set of translated images to optimize an adversarial example. By approximation, TI calculates the gradient at the untranslated image $\hat{x}$ and then average all the shifted gradients. This procedure is equivalent to convolving the gradient with a kernel composed of all the weights.

The major difference between our proposed method and the three aforementioned attacks is that {\bf our method does not rely on task-specific loss functions} (e.g. cross-entropy loss or hinge loss). Instead, it focuses on low-level features, which are presumably task-independent and shared across different models. This is especially critical in the scenario, where the attackers do not know the specific tasks of target models. Our evaluation in \S\ref{sec:experiments} demonstrates improved transferability generated by our method across several different real-world CV tasks.
\section{Methodology}
To construct AEs against a target model, we first establish a source model as the surrogate, to which we have access. Conventionally, the source model is established by training with examples labeled by the target model. That is, the inputs are paired with the labels generated from the target model, instead of the ground truth. In this way, the source model mimics the behavior of the target model. When we construct AEs against the source model, they are likely to transfer to the target model due to such connection.

In our framework, although a source model is still required, there is no need for training new models or querying the target model for labels. Instead, a pretrained public model could simply serve as the source model due to the strong transferability of the AEs generated via our approach. For example, in our experiments, we use pretrained VGG-16, Inception-v3 and Resnet-152, which are publicly available, as the source model $f$.
\begin{algorithm}
    \caption{Dispersion reduction attack}
    \textbf{Input:} A classifier $f$, original sample $\mathbf{x}$, feature map at layer $k$; perturbation budget $\epsilon$ \\
    \textbf{Input:} Attack iterations $T$. \\
    \textbf{Output:} An adversarial example $\mathbf{x}'$ with  ${\parallel \mathbf{x}'-\mathbf{x}\parallel}_\infty\leqslant\epsilon$
    \begin{algorithmic}[1]
    \Procedure{Dispersion reduction}{}
    \State $\mathbf{x}'_0 \gets x$
    \For{$t=0$ to $T-1$}
        \State Forward $\mathbf{x}'_t$ and obtain feature map at layer $k$:
        \begin{equation}
        \mathcal{F}_k = f(\mathbf{x}'_t)|_k
        \end{equation}
        \State Compute dispersion of $\mathcal{F}_k$: $g(\mathcal{F}_k)$
        \State Compute its gradient $w.r.t$ the input: $\bigtriangledown_{\mathbf{x}}g(\mathcal{F}_k)$
        \State Update $\mathbf{x}'_t$:
        \begin{equation}
        \mathbf{x}'_t = \mathbf{x}'_t - \bigtriangledown_{\mathbf{x}}g(\mathcal{F}_k)
        \end{equation}
        \State Project $\mathbf{x}'_t$ to the vicinity of $\mathbf{x}$:
         \begin{equation}
        \mathbf{x}'_t = clip(\mathbf{x}'_t, \mathbf{x}-\epsilon, \mathbf{x}+\epsilon)
        \end{equation}     
    \EndFor
    \State \textbf{return} $\mathbf{x}'_t$ 
    \EndProcedure
    \end{algorithmic}
    \label{alg:dr}
\end{algorithm}
With $f$ as the source model, we construct AEs against it. Existing attacks perturb input images along gradient directions $\bigtriangledown_x J$ that depend on the definition of the task-specific loss function $J$, which not only limits their cross-task transferability but also requires ground-truth labels that are not always available. To mitigate these issues, we propose \emph{dispersion reduction} (DR) attack that formally defines the problem of finding an AE as an optimization problem:
\begin{equation}
\vspace{-0.1cm}
\begin{aligned}
&\min_{\mathbf{x}'} g(f(\mathbf{x}',\theta)) \\
s.t.\   &{\parallel \mathbf{x}'-\mathbf{x}\parallel}_\infty\leqslant\epsilon \\
\end{aligned}
\label{eq:DR}
\end{equation}

\noindent where $f(\cdot)$ is a DNN classifier with output of intermediate feature map, and $g(\cdot)$ calculates the dispersion. Our proposed DR attack, detailed in Algorithm~\ref{alg:dr}, takes a multi-step approach that creates an AE by iteratively reducing the dispersion of an intermediate feature map at layer $k$. Dispersion describes the extent to which a distribution is stretched or squeezed, and there can be different measures of dispersion, such as the standard deviation, and gini coefficient~\cite{statistics}. In this work, we choose standard deviation as the dispersion metric due to its simplicity, and denote it by $g(\cdot)$.

To explain why reducing dispersion could lead to valid attacks, we propose a similar argument used in \cite{goodfellow2014explaining}. Consider a simplified model where $f(\mathbf{x})=\mathbf{a}=(a1,\dots,a_n)$ is the intermediate feature, and $\mathbf{y}=\mathbf{Wa}$ is an affine transformation of the feature (we omit the constant $\mathbf{b}$ for simplicity), resulting the final output logits $\mathbf{y}=(y_1,\dots, y_k)$. In other words, we decompose a DNN classifier into a feature extractor $f(\cdot)$ and an affine transformation. Suppose the correct class is $c$, the logit $y_c$ of a correctly classified example should be the largest, that is $\mathbf{w}_c^\top\mathbf{a}>>\mathbf{w}_i^\top\mathbf{a}$ for $i\neq c$, where $\mathbf{w}_i$ is the $i$th row of $\mathbf{W}$. This indicates $\mathbf{w}_c$ and $\mathbf{a}$ are highly aligned. 

On the other hand, suppose our attack aims to reduce the standard deviation of the feature $\mathbf{a}$. The corresponding adversarial examples $\mathbf{x}'$ leads to a perturbed feature
\begin{equation}
\begin{aligned}
f(\mathbf{x}') &= \mathbf{a}' \approx \mathbf{a}-\alpha\frac{\partial}{\partial\mathbf{a}}Std(\mathbf{a}) \\
               &=\mathbf{a}-2\alpha(\mathbf{a}-\bar{a}\mathbf{1})/(\sqrt{n-1}Std(\mathbf{a}))
\end{aligned}
\label{eq:drive}
\end{equation}
Where $\alpha$ depicts the magnitude of the perturbation on $\mathbf{a}$, $\bar{a}$ is the average of the entries of $\mathbf{a}$, and $\mathbf{1}$ is a column vector with 1 in each entry. Therefore, the change of the logit $y_c$ due to adversarial perturbation is essentially
\begin{equation}
\begin{aligned}
    \Delta y_c &= -2\alpha(\mathbf{w}_c^\top\mathbf{a}-\mathbf{w}_c^\top\mathbf{1}\bar{a})/(\sqrt{n-1}Std(\mathbf{a}))\\
    &=-2\alpha(\mathbf{w}_c^\top\mathbf{a}-n\bar{w}_c\bar{a})/(\sqrt{n-1}Std(\mathbf{a}))\\
    &=-2\alpha \sqrt{n-1}Cov(\mathbf{w}_c,\mathbf{a})/Std(\mathbf{a})< 0
\end{aligned}
\label{equ:cov}
\end{equation}
If we think each entry of $\mathbf{a}$ and $\mathbf{w}_c$ as samples, the $Cov(\mathbf{w}_c,\mathbf{a})$ corresponds to the empirical covariance of these samples. This suggests that as long as $\mathbf{w}_c$ and $\mathbf{a}$ are aligned, our attack can always reduce the logit of the correct class. Note that $\alpha$ is approximately the product of the magnitude of the perturbation on $\mathbf{x}$ and the sensitivity of $f(\cdot)$, therefore the reduction of the logit could be large if $f(\cdot)$ is sensitive, which is often the case in practice. 

In general, $y_c$ could be any activation that is useful for the task, which may not be classification. As long as $y_c$ is large for natural examples, indicating a certain feature is detected, it is always reduced by our attacks according to the analysis above. Thus, our attack is agnostic to tasks and the choice of loss functions.

\section{Experimental Results}
\label{sec:experiments}

\begin{figure*}[h!]
    \centering
    \begin{minipage}[b]{0.24\textwidth}
        \includegraphics[width=\textwidth]{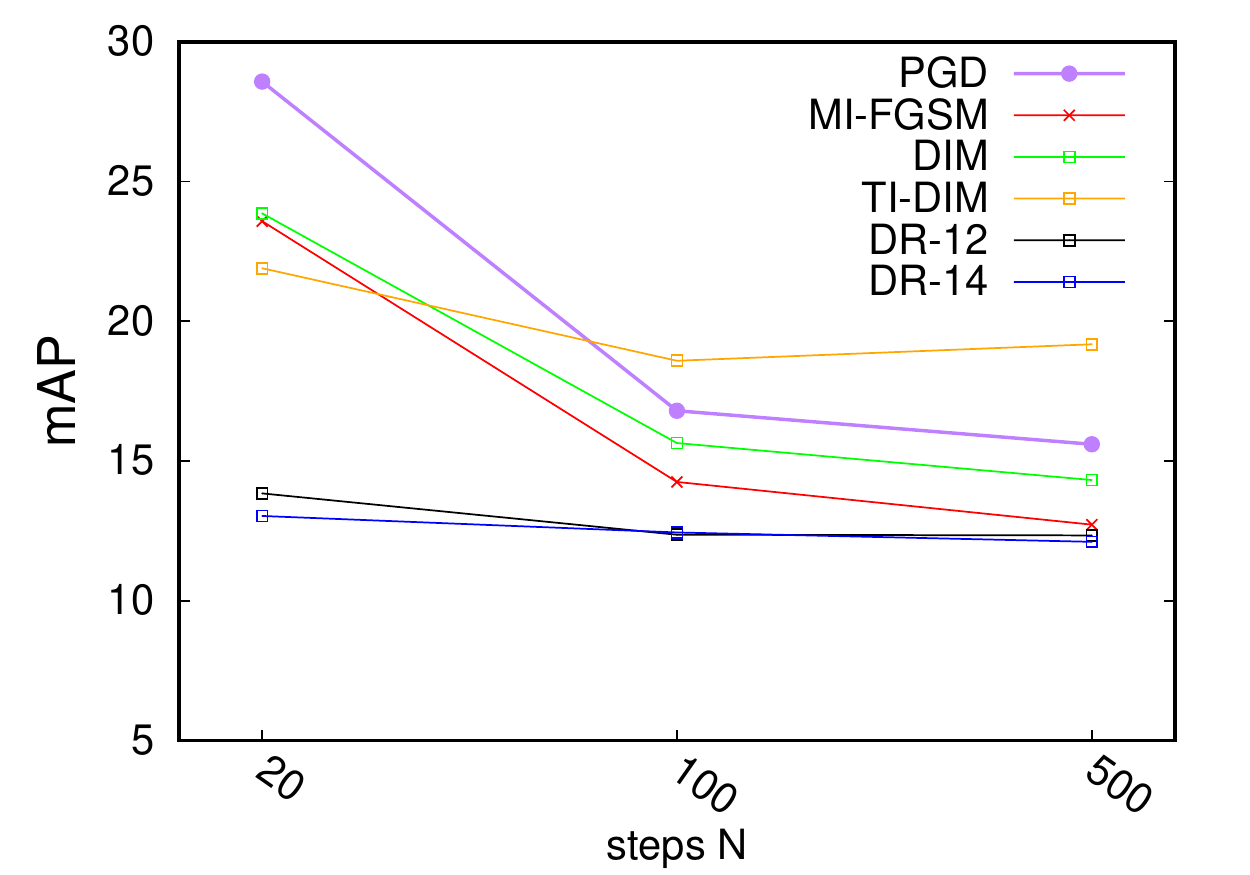}
        \subcaption{SSD-ResNet50}
    \end{minipage}
    \begin{minipage}[b]{0.24\textwidth}
        \includegraphics[width=\textwidth]{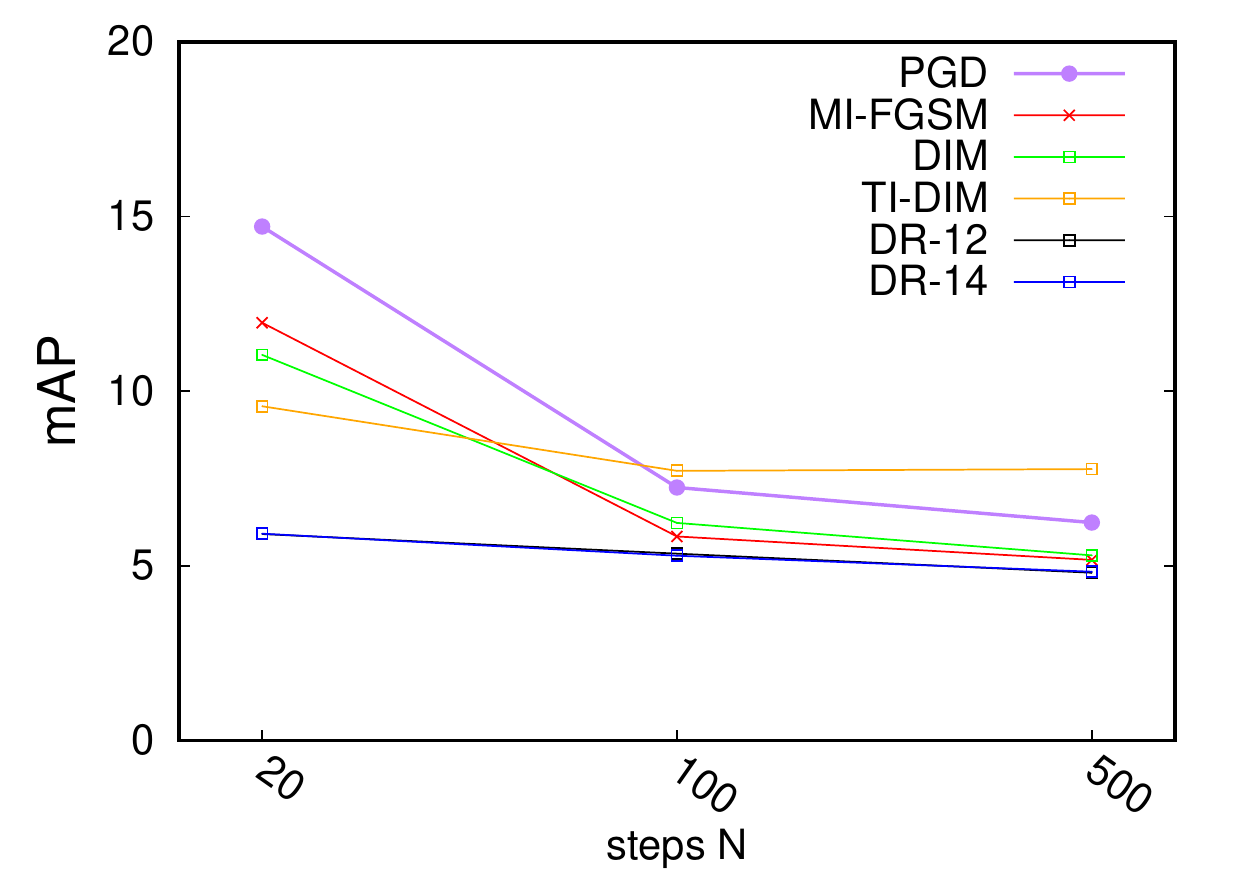}
        \subcaption{RetinaNet-ResNet50}
    \end{minipage}
    \begin{minipage}[b]{0.24\textwidth}
        \includegraphics[width=\textwidth]{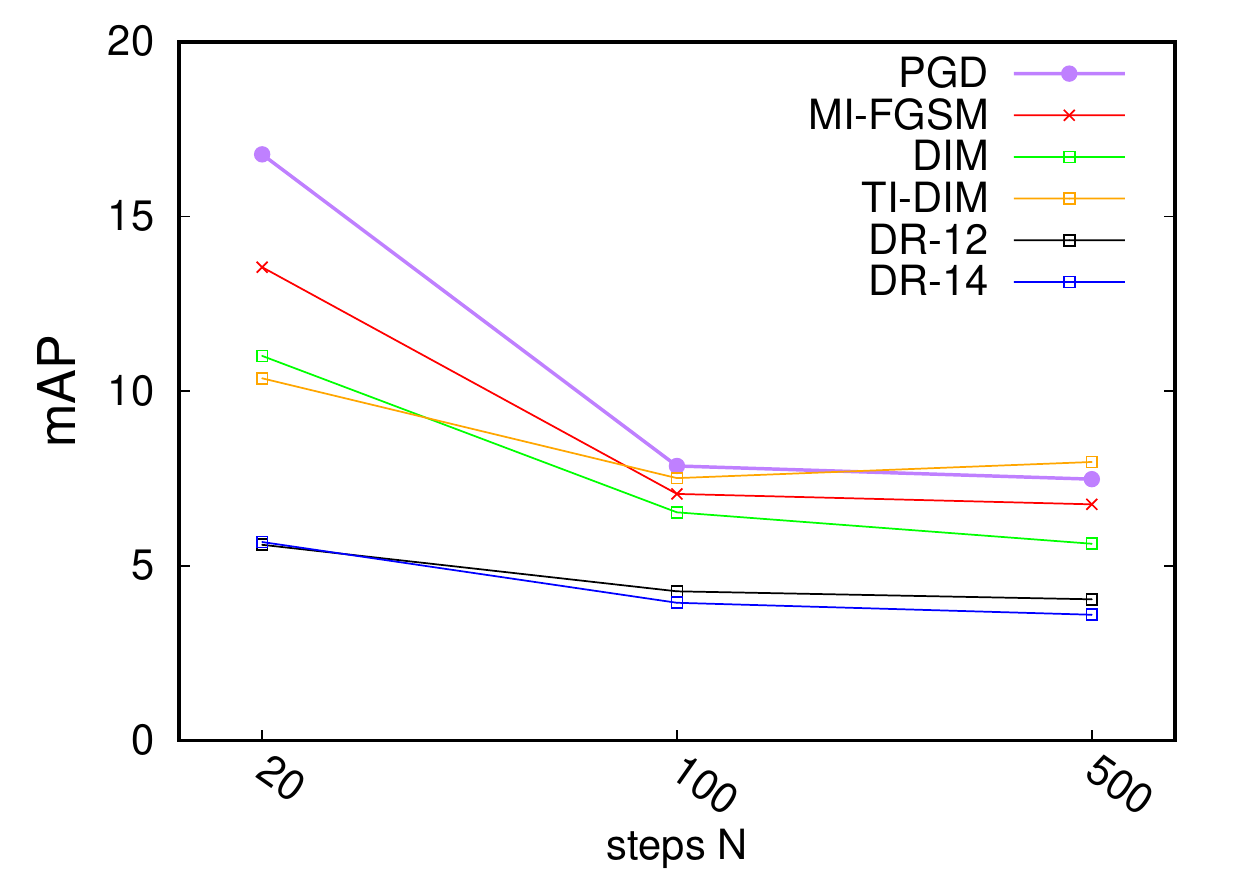}
        \subcaption{SSD-MobileNet}
    \end{minipage}
    \begin{minipage}[b]{0.24\textwidth}
        \includegraphics[width=\textwidth]{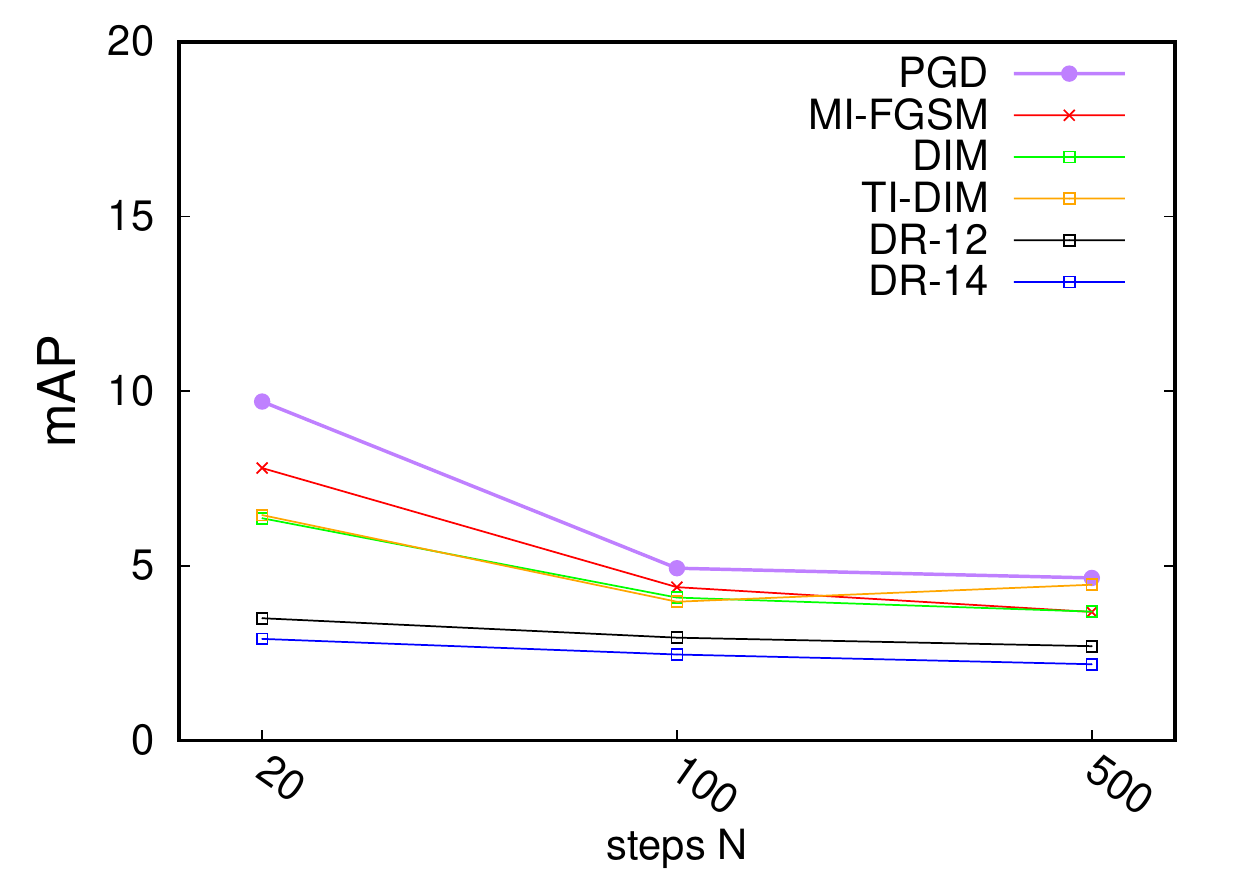}
        \subcaption{FasterRCNN-ResNet50}
    \end{minipage}
    \vspace{-0.3cm}
    \caption{\small{
    \textbf{Results of DR attack with different steps $N$.} We can see that our DR attack outperforms all baselines even starting from small steps (e.g. $N=20$).
    }}
    \label{fig:result_steps}
\end{figure*}

\begin{figure}[hbt!]
    \vspace{0.15cm}
    \centering
    \begin{minipage}[b]{0.23\textwidth}
        \centering
        \includegraphics[width=\textwidth]{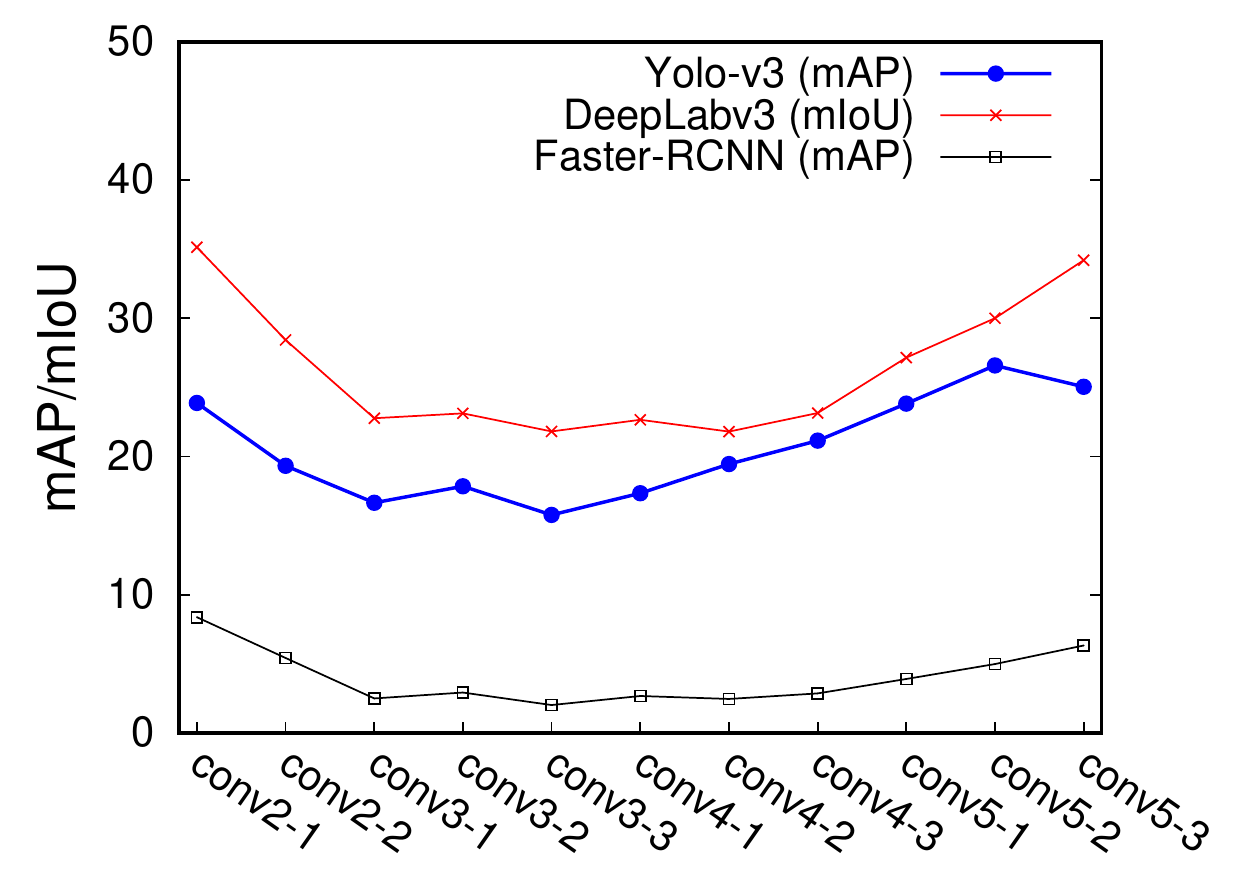}
        \subcaption{mAP/mIoU results.}
        \label{fig:result_layers_result}
    \end{minipage}
    \begin{minipage}[b]{0.23\textwidth}
        \centering
        \includegraphics[width=\textwidth]{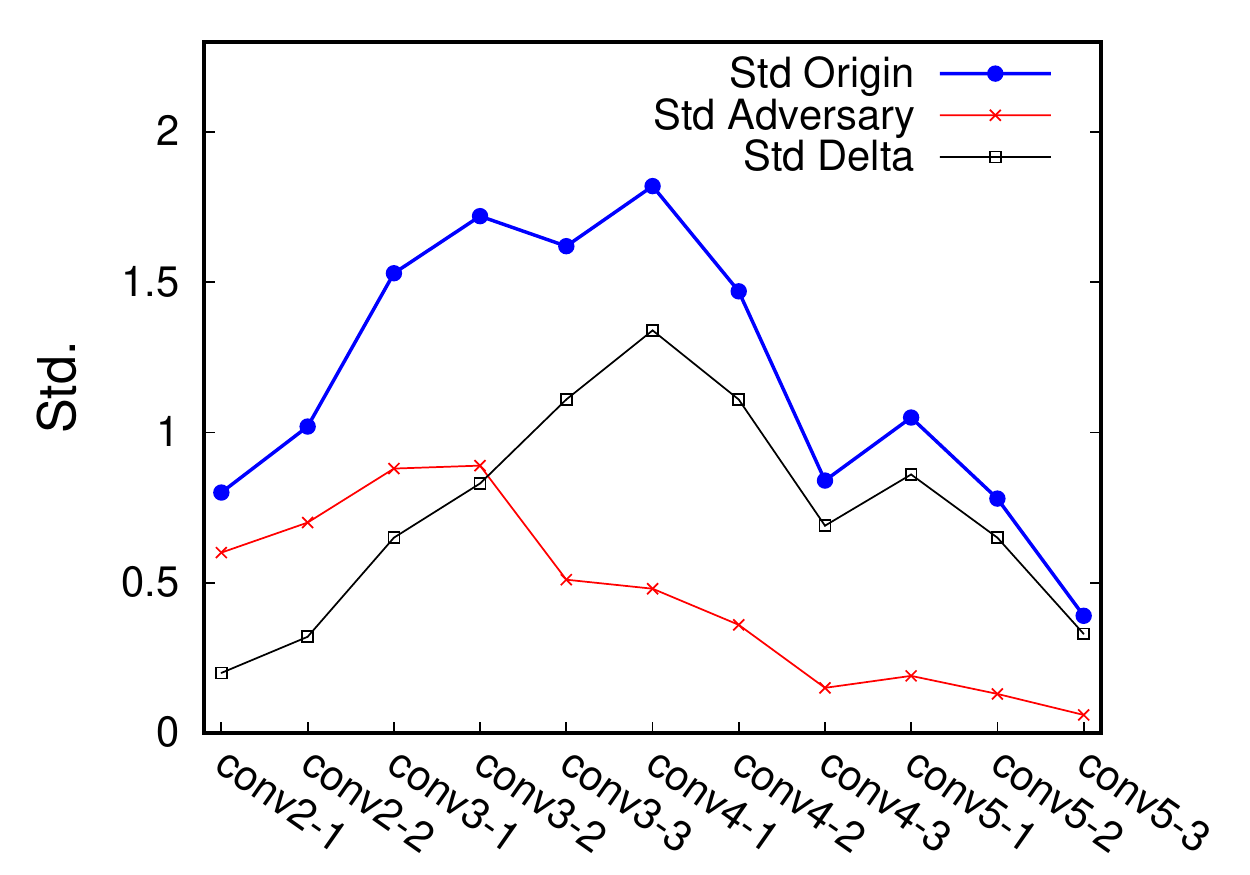}
        \subcaption{Std. before and after attack}
        \label{fig:result_layers_std}
    \end{minipage}
    \vspace{-0.3cm}
    \caption{
    \small{
    \textbf{Results of DR attack with different attack layers of VGG16.} We see that attacking the middle layers results in higher drop in the performance compared to attacking  top  or  bottom  layers.  At  the  same  time,  in the attacking process, the drop in std of middle layers is also larger than the top and bottom layers. This motivates us that we can find a good attack layer by looking at the std drop during the attack.
    }}
    \label{fig:result_layers}
\vspace{-0.5cm}
\end{figure}

We compare our proposed DR attack with the state-of-the-art black-box adversarial attacks on object detection and semantic segmentation tasks (using publicly available models), and commercially deployed Google Cloud Vision (GCV) tasks.
\subsection{Experimental Settings}
\textbf{Network Types:} We consider Yolov3-DarkNet53~\cite{yolov3}, RetinaNet-ResNet50~\cite{retinanet}, SSD-MobileNetv2~\cite{SSD}, Faster R-CNN-ResNet50~\cite{fasterrcnn}, Mask R-CNN-ResNet50~\cite{maskrcnn} as the target object detection models and DeepLabv3Plus-ResNet101~\cite{deeplabv3plus2018}, DeepLabv3-ResNet101~\cite{deeplabv3}, FCN-ResNet101~\cite{fcn} as the target semantic segmentation models. All network models are publicly available, and details are provided in the Appendix. The source networks for generating adversarial examples are VGG16, Inception-v3 and Resnet152 with output image sizes of $(224\times 224)$, $(299\times 299)$ and $(224\times 224)$, respectively.~For the evaluation on COCO2017 and PASCAL VOC2012 datasets, the mAP and mIoU are calculated as the evaluation metrics for detection and semantic segmentation, respectively. Due to the mismatch of different models being trained with different labeling systems (COCO / VOC), only 20 classes that correspond to VOC labels are chosen from COCO labels if a COCO pretrained model is tested on the PASCAL VOC dataset or a VOC pretrained model is tested on the COCO dataset. For the evaluation on ImageNet, since not all test images have the ground truth bounding boxes and pixelwise labels, the mAP and mIoU are calculated as the difference between the outputs of benign / clean images and adversarial images.

\textbf{Implementation details:} We compare our proposed method with projected gradient descent (PGD)~\cite{pgd}, momentum iterative fast gradient sign method (MI-FGSM)~\cite{mifgsm}, diverse inputs method (DIM)~\cite{DIM} and translation-invariant attacks (TI)~\cite{TI}.
As for the hyper-parameters, the maximum perturbation is set to be $\epsilon=16$ for all the experiments with pixel values in [0, 255]. For the proposed DR attacks, the step size $\alpha=4$, and the number of training steps $N=100$. For the baseline methods, we first follow the default settings in \cite{DIM} and \cite{TI} with $\alpha=1$ and $N=20$ for PGD, MI-FGSM and DIM, $\alpha=1.6$ and $N=20$ for TI-DIM. Then, we apply the same hyper-parameters ($\alpha=4$, $N=100$) used with the proposed method to all the baseline methods. For MI-FGSM, we adopt the default decay factor $\mu=1.0$. For DIM and TI-DIM, the transformation probability is set to $p=0.5$.

\subsection{Diagnostics}\label{ssec:diagnostics}
\subsubsection{The effect of training steps $N$}
\label{sssec:effect_steps}
We show the results of attacking SSD-ResNet50, RetinaNet-ResNet50, SSD-MobileNet and Faster RCNN-ResNet50 with different number of training steps ($N=\{20, 100, 500\}$) based on MS COCO2017 validation set. We also compare the proposed DR attack with multiple baselines, namely PGD, MI-FGSM, DIM, TI-DIM. The results are shown in Fig.~\ref{fig:result_steps}. In contrast to the classification-based transfer attacks~\cite{dong2018boosting,xie2018improving,dong2019evading}, we do not observe over-fitting in cross-task transfer attacks for all the tested methods. Therefore, instead of using $N=20$, which is the value used by the baseline attacks we compare with, we can employ larger training steps (N=100), and achieve better attacking performance at the same time. In addition, we can see that our DR attack outperforms all the state-of-the-art baselines for all the step size settings. It should be noticed that DR attack is able to achieve promising results at $N=20$, and the results from the DR attack, using 20 steps, are better than those of baseline methods using 500 steps. This shows that our proposed DR attack has higher efficiency than the baselines.

\begin{table*}[t]
\small
\centering
\begin{tabular}{cl|ccccc}
\hline
\multicolumn{2}{l|}{Detection Results Using Val. Images of} &                                                   \multicolumn{1}{c}{\begin{tabular}[c]{@{}c@{}}Yolov3\\ DrkNet\end{tabular}} & \multicolumn{1}{c}{\begin{tabular}[c]{@{}c@{}}RetinaNet\\ ResNet50\end{tabular}} & \multicolumn{1}{c}{\begin{tabular}[c]{@{}c@{}}SSD\\ MobileNet\end{tabular}} & \multicolumn{1}{c}{\begin{tabular}[c]{@{}c@{}}Faster-RCNN\\ ResNet50\end{tabular}} & \multicolumn{1}{c}{\begin{tabular}[c]{@{}c@{}}Mask-RCNN\\ ResNet50\end{tabular}} \\
\multicolumn{2}{l|}{COCO and VOC Datasets}                                                  & \multicolumn{1}{c}{mAP}                                                     & \multicolumn{1}{c}{mAP}                                                          & \multicolumn{1}{c}{mAP}                                                     & \multicolumn{1}{c}{mAP}                                                            & \multicolumn{1}{c}{mAP}                                                          \\ \hline
\multicolumn{1}{l}{} &                                                  & \multicolumn{1}{c}{COCO/VOC}                                                     & \multicolumn{1}{c}{COCO/VOC}                                                          & \multicolumn{1}{c}{COCO/VOC}                                                     & \multicolumn{1}{c}{COCO/VOC}                                                            & \multicolumn{1}{c}{COCO/VOC}                                                          \\ \hline

VGG16                & PGD ($\alpha$=1, N=20)                & 33.5 / 54.8                                                                   & 14.7 / 31.8                                                                        & 16.8 / 35.9                                                                   & 9.7 / 14.2                                                                           & 10.3 / 15.9                                                                        \\
                     & PGD ($\alpha$=4, N=100)               & 21.6 / 38.7                                                                   & 7.2 / 14.6                                                                         & 7.9 / 18.2                                                                    & 4.9 / 6.4                                                                            & 5.7 / 9.7                                                                          \\
                     & MI-FGSM ($\alpha$=1, N=20)            & 28.4 / 48.9                                                                   & 12.0 / 23.6                                                                        & 13.6 / 29.6                                                                   & 7.8 / 10.9                                                                           & 8.2 / 12.0                                                                         \\
                     & MI-FGSM ($\alpha$=4, N=100)           & \textbf{19.0} / 35.0                                                                   & 5.8 / 10.6                                                                         & 7.0 / 19.1                                                                    & 4.4 / 5.0                                                                            & 4.8 / 7.1                                                                          \\
                     & DIM ($\alpha$=1, N=20)                & 26.7 / 46.9                                                                   & 11.0 / 21.9                                                                        & 11.0 / 22.9                                                                   & 6.4 / 8.2                                                                            & 7.2 / 11.6                                                                         \\
                     & DIM ($\alpha$=4, N=100)               & 20.0 / 37.6                                                                   & 6.2 / 13.0                                                                         & 6.5 / 14.9                                                                    & 4.1 / 5.0                                                                            & 4.6 / 6.7                                                                          \\
                     & TI-DIM ($\alpha$=1.6, N=20)           & 25.8 / 41.4                                                                   & 9.6 / 17.4                                                                         & 10.4 / 19.9                                                                   & 6.5 / 7.5                                                                            & 7.4 / 9.2                                                                          \\
                     & TI-DIM ($\alpha$=4, N=100)            & 19.5 / \textbf{33.4}                                                                   & 7.7 / 13.1                                                                         & 7.5 / 16.7                                                                    & 4.0 / 5.2                                                                            & 4.8 / 6.6                                                                          \\
                     & \textbf{DR} ($\alpha$=4, N=100)\textbf{(ours)} & 19.8 / 38.2                                                                   & \textbf{5.3} / \textbf{8.7}                                                                          & \textbf{3.9} / \textbf{8.2}                                                                     & \textbf{2.5} / \textbf{2.8}                                                                            & \textbf{3.2} / \textbf{5.1}                                                                          \\ \hline
InceptionV3          & PGD ($\alpha$=1, N=20)                & 46.8 / 67.5                                                                   & 23.9 / 51.8                                                                        & 25.2 / 47.4                                                                   & 27.0 / 45.7                                                                          & 27.5 / 48.7                                                                        \\
                     & PGD ($\alpha$=4, N=100)               & 35.3 / 57.1                                                                   & 15.0 / 33.0                                                                       & 14.0 / 31.6                                                                   & 18.2 / 31.7                                                                          & 19.4 / 34.8                                                                        \\
                     & MI-FGSM ($\alpha$=1, N=20)            & 42.0 / 63.9                                                                   & 20.0 / 44.3                                                                        & 20.9 / 43.5                                                                   & 22.8 / 39.3                                                                          & 23.7 / 42.9                                                                        \\
                     & MI-FGSM ($\alpha$=4, N=100)           & 32.4 / 54.0                                                                   & 12.5 / 27.1                                                                        & 13.1 / 29.2                                                                   & 16.3 / 26.9                                                                          & 17.9 / 30.5                                                                        \\
                     & DIM ($\alpha$=1, N=20)                & 32.5 / 54.5                                                                   & 12.9 / 27.5                                                                        & 13.9 / 29.7                                                                   & 14.2 / 24.0                                                                          & 16.3 / 27.7                                                                        \\
                     & DIM ($\alpha$=4, N=100)               & 29.1 / 48.3                                                                   & 10.4 / 20.5                                                                        & 10.4 / 22.0                                                                   & 12.2 / 18.2                                                                          & 13.8 / 44.6                                                                        \\
                     & TI-DIM ($\alpha$=1.6, N=20)           & 32.1 / 50.2                                                                   & 12.8 / 25.8                                                                        & 13.5 / 28.0                                                                   & 12.5 / 20.4                                                                          & 14.4 / 23.0                                                                        \\
                     & TI-DIM ($\alpha$=4, N=100)            & 27.1 / \textbf{42.2}                                                                   & 11.0 / 19.8                                                                        & 10.4 / 22.1                                                                   & 9.9 / 14.6                                                                           & 11.1 / 17.5                                                                        \\
                     & \textbf{DR} ($\alpha$=4, N=100)\textbf{(ours)} & \textbf{24.2} / 45.1                                                                   & \textbf{8.5} / \textbf{18.9}                                                                         & \textbf{9.0} / \textbf{19.5}                                                                    & \textbf{8.3} / \textbf{14.3}                                                                           & \textbf{9.8} / \textbf{17.0}                                                                         \\ \hline
Resnet152            & PGD ($\alpha$=1, N=20)                & 39.4 / 62.0                                                                   & 19.1 / 42.9                                                                        & 19.9 / 41.6                                                                   & 13.8 / 19.4                                                                          & 15.0 / 22.0                                                                        \\
                     & PGD ($\alpha$=4, N=100)               & 28.8 / 51.5                                                                   & 12.2 / 25.9                                                                        & 11.2 / 24.4                                                                   & 8.2 / 11.3                                                                           & 8.8 / 13.9                                                                         \\
                     & MI-FGSM ($\alpha$=1, N=20)            & 35.1 / 58.1                                                                   & 15.8 / 36.2                                                                        & 16.7 / 35.8                                                                   & 11.1 / 16.3                                                                          & 12.2 / 18.1                                                                        \\
                     & MI-FGSM ($\alpha$=4, N=100)           & 26.4 / 48.2                                                                   & 11.2 / 23.5                                                                        & 9.9 / 21.3                                                                    & 7.0 / 9.5                                                                            & 8.2 / 11.4                                                                         \\
                     & DIM ($\alpha$=1, N=20)                & 28.1 / 50.3                                                                   & 12.2 / 26.3                                                                        & 11.0 / 23.9                                                                   & 7.0 / 10.6                                                                           & 7.9 / 12.6                                                                         \\
                     & DIM ($\alpha$=4, N=100)               & 24.7 / 43.2                                                                   & 8.8 / 19.4                                                                         & 7.8 / 16.1                                                                    & 5.1 / 7.1                                                                            & 6.2 / 10.3                                                                         \\
                     & TI-DIM ($\alpha$=1.6, N=20)           & 27.9 / 45.6                                                                   & 11.7 / 21.7                                                                        & 11.3 / 22.5                                                                   & 6.8 / 8.7                                                                            & 7.5 / 9.9                                                                          \\
                     & TI-DIM ($\alpha$=4, N=100)            & \textbf{22.3} / \textbf{36.7}                                                                   & 9.0 / 15.8                                                                         & 8.7 / 19.1                                                                    & 5.0 / 6.6                                                                            & 5.7 / 8.2                                                                          \\
                     & \textbf{DR} ($\alpha$=4, N=100)\textbf{(ours)} & 22.7 / 43.8                                                                   & \textbf{6.8} / \textbf{12.4}                                                                         & \textbf{4.7} / \textbf{7.6}                                                                     & \textbf{2.3} / \textbf{2.8}                                                                            & \textbf{3.0} / \textbf{4.5}                      \\ \hline
\end{tabular}
\vspace{-0.2cm}
\caption{
\small{
\textbf{Detection results using validation images of COCO2017 and VOC2012 datasets.} Our proposed DR attack performs best on 25 out of 30 different cases and achieves 12.8 mAP on average over all the experiments. It creates 3.9 more drop in mAP compared to the best of the baselines (TI-DIM: 16.7 mAP).
}}
\label{tab:result_gt_det}
\end{table*}

\subsubsection{The effect of attack layer}
\label{sssec:effect_attacklayer}
We show the results of attacking different convolutional layers of the VGG16 network with the proposed DR attack based on PASCAL VOC2012 validation set. Fig.~\ref{fig:result_layers_result}, shows the mAP for Yolov3 and faster RCNN, and mIoU for Deeplabv3 and FCN. In Fig.~\ref{fig:result_layers_std}, we plot the standard deviation (std) values before and after the DR attack together with the change. As can be seen, attacking the middle layers of VGG16 results in higher drop in the performance compared to attacking top or bottom layers.~At the same time, the change in std for middle layers is larger compared to the top and bottom layers. We can infer that for initial layers, the budget $\epsilon$ constrains the loss function to reduce the std, while for the layers near the output, the std is already relatively small, and cannot be reduced too much further. Based on this observation, we choose one of the middle layers as the target of the DR attack. More specifically, we attack conv3-3 for VGG16, the last layer of $group-A$ for inception-v3 and the last layer of 2nd group of bottlenecks(conv3-8-3) for ResNet152 in the following experiments.

\subsection{Open Source Model Experiments}
\label{ssec:open_source_model_experiments}
We compare the proposed DR attack with the state-of-the-art adversarial techniques to demonstrate the transferability of our method on public object detection and semantic segmentation models. We use validation sets of ImageNet, VOC2012 and COCO2017 for testing object detection and semantic segmentation tasks. For ImageNet, 5000 correctly classified images from the validation set are chosen. For VOC and COCO, 1000 images from the validation set are chosen. The test images are shared in github repository: \href{https://github.com/anonymous0120/dispersion_reduction_test_images}{dispersion\_reduction\_test\_images}~\cite{our-test-img}.

The results for detection and segmentation on COCO and VOC datasets are shown in Table~\ref{tab:result_gt_det} and Table~\ref{tab:result_gt_seg}, respectively. The results for detection and segmentation on the ImageNet dataset are provided in the Appendix. We also include the table for average results over all the datasets, including the ImageNet, in the Appendix.

As can be seen from Tables~\ref{tab:result_gt_det} and \ref{tab:result_gt_seg}, our proposed method (\textbf{DR}) achieves the best results on 36 out of 42 set of experiments by degrading the performance of the target model by a larger margin. For detection experiments, the \textbf{DR} attack performs best on 25 out of 30 different cases and for semantic segmentation 11 out of 12 different cases. For detection, our proposed attack achieves 12.8 mAP on average over all the experiments. It creates 3.9 more drop in mAP compared to the best of the baselines (TI-DIM: 16.7 mAP). For semantic segmentation, our proposed attack achieves 20.0 mIoU on average over all the experiments. It achieves 5.9 more drop in mIoU compared to the best of the baselines (DIM: 25.9 mIoU).

To summarize the results on the ImageNet dataset provided in the Appendix, our proposed method (\textbf{DR}) achieves the best results in 17 out of 21 sets of experiments. For detection, our proposed attack achieves 7.4 relative-mAP on average over all the experiments. It creates 3.8 more drop in relative-mAP compared to the best of the baselines (TI-DIM: 11.2). For semantic segmentation, our proposed attack achieves 16.9 relative-mIoU on average over all the experiments. It achieves 4.8 more drop in relative-mIoU compared to the best of the baselines (TI-DIM: 21.7).

\begin{table}[ht]
\footnotesize
\begin{tabular}{cl|cc}
\hline
\multicolumn{2}{l|}{Seg. Results Using Val. Images of}            & \multicolumn{1}{c}{\begin{tabular}[c]{@{}c@{}}DeepLabv3\\ ResNet-101\end{tabular}} & \multicolumn{1}{c}{\begin{tabular}[c]{@{}c@{}}FCN\\ ResNet-101\end{tabular}} \\
\multicolumn{2}{l|}{COCO and VOC Datasets}                         & \multicolumn{1}{c}{mIoU}                                                           & \multicolumn{1}{c}{mIoU}                                                     \\ \hline
\multicolumn{1}{l}{} &                         & \multicolumn{1}{c}{COCO/VOC}                                                           & \multicolumn{1}{c}{COCO/VOC}                                                     \\ \hline
VGG16                & PGD ($\alpha$=1, N=20)       & 37.8 / 42.6                                                                          & 26.7 / 29.1                                                                    \\
                     & PGD ($\alpha$=4, N=100)      & 22.3 / 24.0                                                                          & 17.1 / 18.1                                                                    \\
                     & MI-FGSM ($\alpha$=1, N=20)   & 32.8 / 36.2                                                                          & 22.7 / 25.0                                                                    \\
                     & MI-FGSM ($\alpha$=4, N=100)  & 19.9 / 21.6                                                                          & 22.0 / 16.5                                                                    \\
                     & DIM ($\alpha$=1, N=20)       & 30.3 / 33.2                                                                          & 15.5 / 22.4                                                                    \\
                     & DIM ($\alpha$=4, N=100)      & 21.2 / 23.7                                                                          & 16.2 / 16.9                                                                    \\
                     & TI-DIM ($\alpha$=1.6, N=20)  & 29.9 / 31.1                                                                          & 21.9 / 23.0                                                                    \\
                     & TI-DIM ($\alpha$=4, N=100)   & 23.8 / 24.7                                                                          & 18.9 / 19.2                                                                    \\
                     & \textbf{DR} ($\alpha$=4, N=100)\textbf{(ours)} & \textbf{17.2} / \textbf{21.8}                                                                          & \textbf{12.9} / \textbf{14.4}                                                                    \\ \hline
IncV3          & PGD ($\alpha$=1, N=20)       & 49.4 / 56.0                                                                          & 36.8 / 40.1                                                                    \\
                     & PGD ($\alpha$=4, N=100)      & 37.1 / 41.3                                                                          & 26.1 / 28.3                                                                    \\
                     & MI-FGSM ($\alpha$=1, N=20)   & 44.2 / 51.1                                                                          & 32.4 / 35.4                                                                    \\
                     & MI-FGSM ($\alpha$=4, N=100)  & 33.7 / 39.1                                                                          & 24.0 / 35.4                                                                    \\
                     & DIM ($\alpha$=1, N=20)       & 35.7 / 40.4                                                                          & 24.9 / 27.2                                                                    \\
                     & DIM ($\alpha$=4, N=100)      & 30.4 / 33.9                                                                          & 21.3 / 22.3                                                                    \\
                     & TI-DIM ($\alpha$=1.6, N=20)  & 35.3 / 37.0                                                                          & 26.4 / 27.7                                                                    \\
                     & TI-DIM ($\alpha$=4, N=100)   & 29.0 / 29.8                                                                          & 22.5 / 23.5                                                                    \\
                     & \textbf{DR} ($\alpha$=4, N=100)\textbf{(ours)} & \textbf{23.2} / \textbf{29.2}                                                                          & \textbf{17.1} / \textbf{20.9}                                                                    \\ \hline
Res152            & PGD ($\alpha$=1, N=20)       & 45.2 / 50.2                                                                          & 30.7 / 34.6                                                                    \\
                     & PGD ($\alpha$=4, N=100)      & 31.5 / 35.1                                                                          & 21.6 / 24.0                                                                    \\
                     & MI-FGSM ($\alpha$=1, N=20)   & 39.9 / 43.9                                                                          & 26.4 / 29.9                                                                    \\
                     & MI-FGSM ($\alpha$=4, N=100)  & 28.2 / 32.2                                                                          & 19.9 / 22.1                                                                    \\
                     & DIM ($\alpha$=1, N=20)       & 31.3 / 35.5                                                                          & 22.3 / 23.9                                                                    \\
                     & DIM ($\alpha$=4, N=100)      & 25.9 / 28.8                                                                          & 19.0 / 19.9                                                                    \\
                     & TI-DIM ($\alpha$=1.6, N=20)  & 31.8 / 33.9                                                                          & 23.7 / 25.2                                                                    \\
                     & TI-DIM ($\alpha$=4, N=100)   & 26.6 / \textbf{26.6}                                                                          & 20.3 / 21.4                                                                    \\
                     & \textbf{DR} ($\alpha$=4, N=100)\textbf{(ours)} & \textbf{22.7} / 27.0                                                                          & \textbf{16.4} / \textbf{17.6}                                                                    \\ \hline
\end{tabular}
\vspace{-0.3cm}
\caption{
\small{
\textbf{Semantic Segmentation results using validation images of COCO2017 and VOC2012 datasets.} Our proposed DR attack performs best on 11 out of 12 different cases and achieves 20.0 mIoU on average over all the experiments. It achieves 5.9 more drop in mIoU compared to the best of the baselines (DIM: 25.9 mIoU).
}}
\label{tab:result_gt_seg}
\end{table}

\subsection{Cloud API Experiments}
\label{ssec:API_experiments}
We compare proposed DR attack with the state-of-the-art adversarial techniques to enhance transferability on commercially deployed Google Cloud Vision (GCV) tasks~\footnote{https://cloud.google.com/vision/docs}:
\begin{itemize}
    \item Image Label Detection (\textbf{\texttt{Labels}}) classifies image into broad sets of categories.
    \item Object Detection (\textbf{\texttt{Objects}}) detects multiple objects with their labels and bounding boxes in an image.
    \item Image Texts Recognition (\textbf{\texttt{Texts}}) detects and recognize text within an image, which returns their bounding boxes and transcript texts.
    \item Explicit Content Detection (\textbf{\texttt{SafeSearch}}) detects explicit content such as adult or violent content within an image, and returns the likelihood.
\end{itemize}

\begin{figure}
\centering
\includegraphics[width=\columnwidth]{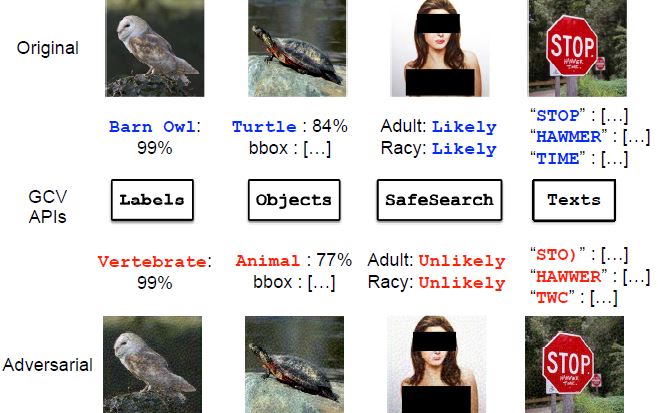}
\vspace{-0.7cm}
\caption{
\small{
Visualization of images chosen from testing set and their corresponding AEs generated by DR. All the AEs are generated on VGG-16 \texttt{conv3.3} layer, with perturbations clipped by $l_\infty\leq16$, and they effectively fool the four GCV APIs as indicated by their outputs.}}
\label{fig:demo}
\end{figure}

\textbf{Datasets.} We use ImageNet validation set for testing \texttt{Labels} and \texttt{Objects}, and the NSFW Data Scraper~\cite{nsfw_dataset} and COCO-Text~\cite{icdar} dataset for evaluating against \texttt{SafeSearch} and \texttt{Texts}, respectively. We randomly choose 100 images from each dataset for our evaluation, and  Fig.~\ref{fig:demo} shows sample images in our test set. Please note that due to the API query fees, larger scale experiments could not be performed for this part.

\begin{table*}[ht]
\small
\centering
\begin{threeparttable}[b]
\begin{tabular}{|c|c|ccccc|}
\hline
\multirow{2}{*}{Model} & \multirow{2}{*}{Attack} & \multicolumn{1}{c}{\textbf{\texttt{Labels}}} & \multicolumn{1}{c}{\textbf{\texttt{Objects}}} & \multicolumn{1}{c}{\textbf{\texttt{SafeSearch}}} & \multicolumn{2}{c|}{
\textbf{\texttt{Texts}}} \\ \cline{3-7}
 &  & \multicolumn{1}{c}{acc.} & \multicolumn{1}{c}{mAP (IoU=0.5)} & \multicolumn{1}{c}{acc.} & \multicolumn{1}{c}{AP (IoU=0.5)} & \multicolumn{1}{c|}{C.R.W\tnote{2}} \\ \hline \hline
\multicolumn{2}{|c|}{baseline (SOTA)\tnote{1}} & 82.5\% & 73.2 & 100\% & 69.2 & 76.1\% \\ \hline \hline
\multirow{3}{*}{VGG-16} & MI-FGSM & 41\% & 42.6 & 62\% & 38.2 & 15.9\% \\
 & DIM & 39\% & 36.5 & 57\% & 29.9 & 16.1\% \\
 & DR (\textbf{Ours}) & \textbf{23}\% & \textbf{32.9} & \textbf{35}\% & \textbf{20.9} & \textbf{4.1}\% \\ \hline \hline
\multirow{3}{*}{Resnet-152} & MI-FGSM & 37\% & 41.0 & 61\% & 40.4 & 17.4\% \\
 & DIM & 49\% & 46.7 & 60\% & \textbf{34.2} & 15.1\% \\
 & DR (\textbf{Ours}) & \textbf{25}\% & \textbf{33.3} & \textbf{31}\% & 34.6 & \textbf{9.5}\% \\ \hline
\end{tabular}
\begin{tablenotes}
\small
    \item[1] The baseline performance of GCV models cannot be measured due to the mismatch between original labels and labels used by Google. We use the GCV prediction results on original images as ground truth, thus the baseline performance should be 100\% for all accuracy and 100.0 for mAP and AP. Here we provide state-of-the-art performance~\cite{ILSVRC2017, keras, icdar, nsfw_dataset} for reference.
    \item[2] Correctly recognized words (C.R.W)~\cite{icdar}.
\end{tablenotes}
\vspace{-0.2cm}
\caption{
\small
\textbf{The degraded performance of four Google Cloud Vision models, where we attack a single model from the left column.} Our proposed DR attack degrades the accuracy of \textbf{\texttt{Lables}} and \texttt{\textbf{SafeSearch}} to 23\% and 35\%, the mAP of \textbf{\texttt{Objects}} and \textbf{\texttt{Texts}} to 32.9 and 20.9, the word recognition accuracy of \textbf{\texttt{Texts}} to only 4.1\%, which outperform existing attacks.
}\label{table:api_results}
\end{threeparttable}
\end{table*}

\textbf{Experiment setup.} To generate the AEs, We use normally trained VGG-16 and Resnet-152 as our source models, since Resnet-152 is commonly used by MI-FGSM and DIM for generation~\cite{xie2018improving, dong2018boosting}. Since DR attack targets a specific layer, we choose \texttt{conv3.3} for VGG-16 and \texttt{conv3.8.3} for Resnet-152 as per the profiling result in Table~\ref{table:api_results} and discussion in Sec.~\ref{sssec:effect_attacklayer}.

\textbf{Attack parameters.} We follow the default settings in~\cite{dong2018boosting} with the momentum decay factor $\mu=1$ when implementing the MI-FGSM attack. For the DIM attack, we set probability $p=0.5$ for the stochastic transformation function $T(x;p)$ as in~\cite{xie2018improving}, and use the same decay factor $\mu=1$ and total iteration number $N=20$ as in the vanilla MI-FGSM. For our proposed DR attack, we do not rely on FGSM method, and instead use Adam optimizer ($\beta_1=0.98$, $\beta_2=0.99$) with learning rate of $5e^{-2}$ to reduce the dispersion of target feature map. The maximum perturbation of all attacks in the experiments are limited by clipping at $l_\infty=16$, which is still considered less perceptible for human observers~\cite{luo2015foveation}.

\textbf{Evaluation metrics.} We perform adversarial attacks only on single network and test them on the four black-box GCV models. The effectiveness of attacks is measured by the model performance under attacks. As the labels from original datasets are different from labels used by GCV, we use the prediction results of GCV APIs on the original data as the ground truth, which gives a baseline performance of 100\% relative accuracy or 100.0 relative mAP and AP respectively.


\textbf{Results.} We provide the state-of-the-art results on each CV task as reference in Table~\ref{table:api_results}. As shown in Table~\ref{table:api_results}, DR outperforms other baseline attacks by degrading the target model performance by a larger margin. For example, the adversarial examples crafted by DR on VGG-16 model brings down the accuracy of  \textbf{\texttt{Labels}} to only 23\%, and \textbf{\texttt{SafeSearch}} to 35\%. Adversarial examples created with the DR, also degrade mAP of \textbf{\texttt{Objects}} to 32.9\% and AP of text localization to 20.9\%, and with barely 4.1\% accuracy in recognizing words.~Strong baselines like MI-FGSM and DIM, on the other hand, only cause 38\% and 43\% success rate, respectively, when attacking \texttt{SafeSearch}, and are less effective compared with DR when attacking all other GCV models. The results demonstrate the better cross-task transferability of the dispersion reduction attack.


Figure~\ref{fig:demo} shows example of each GCV model's output for original and adversarial examples. The performance of \textbf{\texttt{Labels}} and \textbf{\texttt{SafeSearch}} are measured by the accuracy of classification. More specifically, we use \emph{top\-1} accuracy for \textbf{\texttt{Labels}}, and use the accuracy for detecting the given porn images as \texttt{LIKELY} or \texttt{VERY\_LIKELY} being \texttt{adult} for \textbf{\texttt{SafeSearch}}.
The performance of \textbf{\texttt{Objects}} is given by the mean average precision (mAP) at IoU=0.5. For \textbf{\texttt{Texts}}, we follow the bi-fold evaluation method of ICDAR 2017 Challenge~\cite{icdar}. We measure text localization accuracy using average precision (AP) of bounding boxes at IoU=0.5, and evaluate the word recognition accuracy with correctly recognized words (C.R.W) that are case insensitive.

When comparing the effectiveness of attacks on different generation models, the results that DR generates adversarial examples that transfer better across these four commercial APIs still hold. The visualization in Fig.~\ref{fig:demo} shows that the perturbed images with $l_\infty\leq16$ well maintain their visual similarities with original images, but fool the real-world computer vision systems. 
\section{Discussion and Conclusion} \label{sec:Concl}
In this paper, we propose a \emph{Dispersion Reduction} (DR) attack to improve the cross-task transferability of adversarial examples. Specifically, our method reduces the dispersion of intermediate feature maps by iterations. Compared to existing black-box attacks, the results on MS COCO, PASCAL VOC and ImageNet show that our proposed method performs better on attacking black-box cross-CV-task models.
One intuition behind the DR attack is that by minimizing the dispersion of feature maps, images become "featureless". This is because few features can be detected if neuron activations are suppressed by perturbing the input (Fig.~\ref{fig:illustration}). Moreover, with the observation that low-level features bear more similarities across CV models, we hypothesize that the DR attack would produce transferable adversarial examples when one of the middle convolution layers is targeted. Evaluation on different CV tasks shows that this enhanced attack greatly degrades model performance by a large margin compared to the state-of-the-art attacks, and thus would facilitate evasion attacks against a different task model or even an ensemble of CV-based detection mechanisms. We hope that our proposed attack can serve as benchmark for evaluating robustness of future defense mechanisms. Code is publicly available at: \href{https://github.com/anonymous0120/dr}{https://github.com/anonymous0120/dr}.

\balance
\clearpage
{\small
\bibliographystyle{ieee_fullname}
\bibliography{egbib}
}

\clearpage
\appendix
\section{Target models}
The backbones and datasets of pretrained weights for target models are shown in Table~\ref{tab:target_models}.

\begin{table}[h]
\footnotesize
\centering
\begin{tabular}{lll}
\hline
Models         & Backbone    & Pretrained Dataset     \\ \hline
Yolov3\cite{yolov3}\cite{yolov3-git}         & DarkNet53   & COCO                   \\
RetineNet\cite{retinanet}\cite{retinanet-git}     & ResNet50    & COCO                   \\
SSD\cite{SSD}\cite{SSD-git}           & MobileNet & COCO                   \\
Faster R-CNN\cite{fasterrcnn}\cite{torchvision}   & ResNet50    & COCO                   \\
Mask R-CNN\cite{maskrcnn}\cite{torchvision}     & ResNet50    & COCO                   \\
DeepLabv3\cite{deeplabv3}\cite{torchvision}      & ResNet101   & sub COCO in VOC labels \\
FCN~\cite{fcn}\cite{torchvision}            & ResNet101   & sub COCO in VOC labels \\ \hline
\end{tabular}
\caption{
Backbone and pretrained dataset for target models.
}
\label{tab:target_models}
\end{table}

\section{Experiments on ImageNet}
\label{sec:supp:ImagenetExp}
We have performed adversarial attacks on randomly chosen 5000 correctly classified images from the ImageNet validation set. The accuracies for detection and segmentation are shown in Table~\ref{table:result_imagenet_det} and Table~\ref{table:result_imagenet_seg}, respectively. Since there are no ground truth annotations and masks for the test images, the performance metrics are selected as the relative mAP/mIoU for detection and semantic segmentation respectively. In other words, the predictions from benign samples are regarded as the ground truth and predictions from adversarial examples are regarded as inference results.

Our proposed method (DR) achieves the best results in 17 out of 21 sets of experiments (81.0\%) by degrading the performance of the target model by a larger margin. For detection, our proposed attack reduces the mAP, on average, to 7.41 over all the experiments.  It creates 3.8 more drop in mAP compared to the best of the baselines (TI-DIM: 11.2 mAP). For semantic segmentation, our proposed attack achieves 16.93 mIoU on average over all the experiments. It achieves 4.76 more drop in mIoU compared to the best of the baselines (DIM: 21.69 mIoU).

\begin{table}[h]
\footnotesize
\centering
\begin{tabular}{l|cc}
\hline
\multicolumn{1}{c|}{\multirow{2}{*}{Avg. Res.}} & Det.                & Seg.                 \\
\multicolumn{1}{c|}{}                           & mAP                 & mIoU                 \\ \hline
       &   \footnotesize{COCO\&VOC/ImageNet}       &         \\ \hline
PGD                                             & 26.1 / 19.1         & 33.6 / 28.8          \\
MI-FGSM                                         & 22.8 / 15.6         & 30.6 / 25.2          \\
DIM                                             & 18.6 / 11.5         & 25.9 / 21.8          \\
TI-DIM                                          & 16.7 / 11.2         & 26.4 / 21.7          \\
\textbf{DR (Ours)}                              & \textbf{12.8 / 7.4} & \textbf{20.0 / 16.9} \\ \hline
\end{tabular}
\caption{Average results for detection and segmentation using COCO, VOC and ImageNet validation images.}
\label{table:avg_det_seg}
\end{table}

\begin{table*}[]
\centering
\footnotesize
\begin{tabular}{cl|ccccc}
\hline
                             & \multicolumn{1}{c|}{}   & \multicolumn{1}{c}{\begin{tabular}[c]{@{}c@{}}Yolov3\\ DrkNet\end{tabular}} & \multicolumn{1}{c}{\begin{tabular}[c]{@{}c@{}}RetinaNet\\ ResNet50\end{tabular}} & \multicolumn{1}{c}{\begin{tabular}[c]{@{}c@{}}SSD\\ MobileNet\end{tabular}} & \multicolumn{1}{c}{\begin{tabular}[c]{@{}c@{}}Faster-RCNN\\ ResNet50\end{tabular}} & \multicolumn{1}{c}{\begin{tabular}[c]{@{}c@{}}Mask-RCNN\\ ResNet50\end{tabular}} \\
                             & \multicolumn{1}{c|}{}   & \multicolumn{1}{c}{mAP}                                                     & \multicolumn{1}{c}{mAP}                                                          & \multicolumn{1}{c}{mAP}                                                     & \multicolumn{1}{c}{mAP}                                                            & \multicolumn{1}{c}{mAP}                                                          \\ \hline
\multirow{9}{*}{VGG16}       & PGD($\alpha$=1,N=20)       & 31.6                                                                        & 19.1                                                                             & 19.5                                                                        & 6.4                                                                                & 7.1                                                                              \\
                             & PGD($\alpha$=4,N=100)      & 18.7                                                                        & 7.0                                                                              & 7.7                                                                         & 2.8                                                                                & 3.3                                                                              \\
                             & MI-FGSM($\alpha$=1,N=20)   & 25.9                                                                        & 13.4                                                                             & 15.2                                                                        & 4.7                                                                                & 5.0                                                                              \\
                             & MI-FGSM($\alpha$=4,N=100)  & 16.4                                                                        & 5.0                                                                              & 6.6                                                                         & 1.8                                                                                & 2.2                                                                              \\
                             & DIM($\alpha$=1,N=20)       & 23.4                                                                        & 11.3                                                                             & 11.5                                                                        & 3.7                                                                                & 4.5                                                                              \\
                             & DIM($\alpha$=4,N=100)      & 17.2                                                                        & 5.8                                                                              & 6.3                                                                         & 2.2                                                                                & 2.7                                                                              \\
                             & TI-DIM($\alpha$=1.6,N=20)  & 21.5                                                                        & 10.2                                                                             & 11.6                                                                        & 3.5                                                                                & 4.0                                                                              \\
                             & TI-DIM($\alpha$=4,N=100)   & \textbf{16.3}                                                               & 7.8                                                                              & 8.6                                                                         & 2.3                                                                                & 2.7                                                                              \\
                             & \textbf{DR}($\alpha$=4,N=100)(\textbf{ours}) & 17.0                                                                        & \textbf{3.6}                                                                     & \textbf{4.1}                                                                & \textbf{1.2}                                                                       & \textbf{1.5}                                                                     \\ \hline
\multirow{9}{*}{InceptionV3} & PGD($\alpha$=1,N=20)       & 51.3                                                                        & 36.6                                                                             & 33.9                                                                        & 25.9                                                                               & 25.1                                                                             \\
                             & PGD($\alpha$=4,N=100)      & 33.3                                                                        & 16.4                                                                             & 16.2                                                                        & 14.1                                                                               & 14.7                                                                             \\
                             & MI-FGSM($\alpha$=1,N=20)   & 44.6                                                                        & 27.4                                                                             & 27.5                                                                        & 19.8                                                                               & 20.1                                                                             \\
                             & MI-FGSM($\alpha$=4,N=100)  & 30.3                                                                        & 14.1                                                                             & 15.3                                                                        & 11.9                                                                               & 12.5                                                                             \\
                             & DIM($\alpha$=1,N=20)       & 30.6                                                                        & 15.2                                                                             & 16.4                                                                        & 11.0                                                                               & 11.7                                                                             \\
                             & DIM($\alpha$=4,N=100)      & 25.3                                                                        & 10.2                                                                             & 10.6                                                                        & 6.9                                                                                & 8.2                                                                              \\
                             & TI-DIM($\alpha$=1.6,N=20)  & 30.6                                                                        & 15.4                                                                             & 16.1                                                                        & 9.4                                                                                & 10.3                                                                             \\
                             & TI-DIM($\alpha$=4,N=100)   & 23.7                                                                        & 11.2                                                                             & 12.2                                                                        & 6.8                                                                                & 7.0                                                                              \\
                             & \textbf{DR}($\alpha$=4,N=100)(\textbf{ours}) & \textbf{21.1}                                                               & \textbf{8.6}                                                                     & \textbf{9.4}                                                                & \textbf{4.5}                                                                       & \textbf{5.3}                                                                     \\ \hline
\multirow{9}{*}{Resnet152}   & PGD($\alpha$=1,N=20)       & 40.8                                                                        & 27.6                                                                             & 27.0                                                                        & 10.4                                                                               & 10.8                                                                             \\
                             & PGD($\alpha$=4,N=100)      & 27.2                                                                        & 13.4                                                                             & 13.0                                                                        & 5.0                                                                                & 6.1                                                                              \\
                             & MI-FGSM($\alpha$=1,N=20)   & 33.9                                                                        & 20.3                                                                             & 21.2                                                                        & 7.6                                                                                & 8.0                                                                              \\
                             & MI-FGSM($\alpha$=4,N=100)  & 24.6                                                                        & 11.4                                                                             & 11.8                                                                        & 3.9                                                                                & 4.7                                                                              \\
                             & DIM($\alpha$=1,N=20)       & 26.9                                                                        & 13.2                                                                             & 13.0                                                                        & 4.4                                                                                & 5.3                                                                              \\
                             & DIM($\alpha$=4,N=100)      & 22.2                                                                        & 9.3                                                                              & 8.7                                                                         & 2.9                                                                                & 3.7                                                                              \\
                             & TI-DIM($\alpha$=1.6,N=20)  & 25.3                                                                        & 13.0                                                                             & 13.3                                                                        & 4.2                                                                                & 5.0                                                                              \\
                             & TI-DIM($\alpha$=4,N=100)   & \textbf{19.5}                                                               & 9.4                                                                              & 9.8                                                                         & 2.7                                                                                & 2.9                                                                              \\
                             & \textbf{DR}($\alpha$=4,N=100)(\textbf{ours}) & 21.0                                                                        & \textbf{6.2}                                                                     & \textbf{4.8}                                                                & \textbf{1.3}                                                                       & \textbf{1.6}                                                                     \\ \hline
\end{tabular}
\vspace{-0.2cm}
\caption{
\small
Detection results for ImageNet.}
\label{table:result_imagenet_det}
\end{table*}

\begin{table*}[]
\footnotesize
\centering
\begin{tabular}{cl|cc}
\hline
                             & \multicolumn{1}{c|}{}   & \multicolumn{1}{c}{\begin{tabular}[c]{@{}c@{}}DeepLabv3\\ ResNet101\end{tabular}} & \multicolumn{1}{c}{\begin{tabular}[c]{@{}c@{}}FCN\\ ResNet101\end{tabular}} \\
                             & \multicolumn{1}{c|}{}   & \multicolumn{1}{c}{mIoU}                                                          & \multicolumn{1}{c}{mIoU}                                                    \\ \hline
\multirow{9}{*}{VGG16}       & PGD($\alpha$=1,N=20)       & 30.3                                                                              & 24.6                                                                        \\
                             & PGD($\alpha$=4,N=100)      & 17.5                                                                              & 15.1                                                                        \\
                             & MI-FGSM($\alpha$=1,N=20)   & 25.4                                                                              & 20.8                                                                        \\
                             & MI-FGSM($\alpha$=4,N=100)  & \textbf{15.5}                                                                     & 13.9                                                                        \\
                             & DIM($\alpha$=1,N=20)       & 24.7                                                                              & 19.0                                                                        \\
                             & DIM($\alpha$=4,N=100)      & 17.1                                                                              & 14.5                                                                        \\
                             & TI-DIM($\alpha$=1.6,N=20)  & 23.8                                                                              & 20.0                                                                        \\
                             & TI-DIM($\alpha$=4,N=100)   & 18.3                                                                              & 16.5                                                                        \\
                             & \textbf{DR}($\alpha$=4,N=100)(\textbf{ours}) & 16.5                                                                              & \textbf{12.4}                                                               \\ \hline
\multirow{9}{*}{InceptionV3} & PGD($\alpha$=1,N=20)       & 47.3                                                                              & 37.5                                                                        \\
                             & PGD($\alpha$=4,N=100)      & 31.0                                                                              & 24.4                                                                        \\
                             & MI-FGSM($\alpha$=1,N=20)   & 40.5                                                                              & 31.8                                                                        \\
                             & MI-FGSM($\alpha$=4,N=100)  & 28.3                                                                              & 22.8                                                                        \\
                             & DIM($\alpha$=1,N=20)       & 30.4                                                                              & 24.4                                                                        \\
                             & DIM($\alpha$=4,N=100)      & 25.0                                                                              & 20.0                                                                        \\
                             & TI-DIM($\alpha$=1.6,N=20)  & 28.1                                                                              & 24.4                                                                        \\
                             & TI-DIM($\alpha$=4,N=100)   & 22.1                                                                              & 20.6                                                                        \\
                             & \textbf{DR}($\alpha$=4,N=100)(\textbf{ours}) & \textbf{19.7}                                                                     & \textbf{17.2}                                                               \\ \hline
\multirow{9}{*}{Resnet152}   & PGD($\alpha$=1,N=20)       & 39.5                                                                              & 31.1                                                                        \\
                             & PGD($\alpha$=4,N=100)      & 26.4                                                                              & 20.9                                                                        \\
                             & MI-FGSM($\alpha$=1,N=20)   & 33.5                                                                              & 26.3                                                                        \\
                             & MI-FGSM($\alpha$=4,N=100)  & 24.5                                                                              & 19.3                                                                        \\
                             & DIM($\alpha$=1,N=20)       & 26.8                                                                              & 21.0                                                                        \\
                             & DIM($\alpha$=4,N=100)      & 21.7                                                                              & 17.3                                                                        \\
                             & TI-DIM($\alpha$=1.6,N=20)  & 26.2                                                                              & 21.9                                                                        \\
                             & TI-DIM($\alpha$=4,N=100)   & \textbf{20.1}                                                                     & 18.3                                                                        \\
                             & \textbf{DR}($\alpha$=4,N=100)(\textbf{ours}) & 20.5                                                                              & \textbf{15.3}                                                               \\ \hline
\end{tabular}
\vspace{-0.2cm}
\caption{
\small
Segmentation Results for ImageNet.
}
\label{table:result_imagenet_seg}
\end{table*}

\begin{figure*}[]
    \centering
    \begin{minipage}[b]{0.13\textwidth}
        \includegraphics[width=\textwidth]{det_1_clean}
    \end{minipage}
    \begin{minipage}[b]{0.13\textwidth}
        \includegraphics[width=\textwidth]{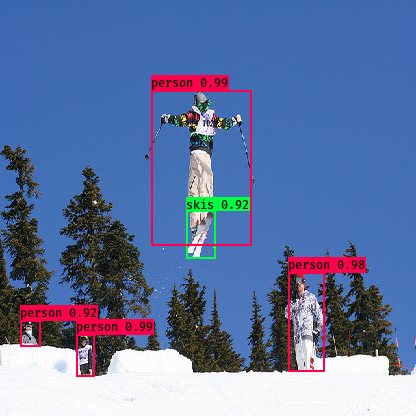}
    \end{minipage}
    \begin{minipage}[b]{0.13\textwidth}
        \includegraphics[width=\textwidth]{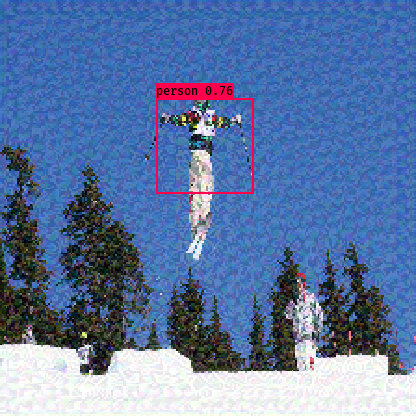}
    \end{minipage}
    \begin{minipage}[b]{0.13\textwidth}
        \includegraphics[width=\textwidth]{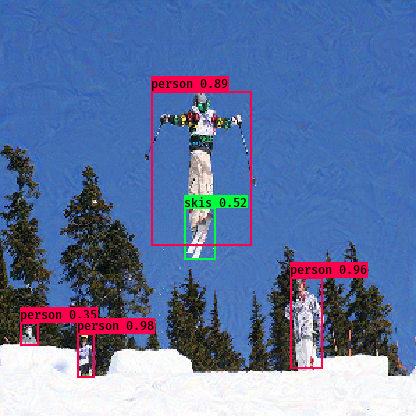}
    \end{minipage}
    \begin{minipage}[b]{0.13\textwidth}
        \includegraphics[width=\textwidth]{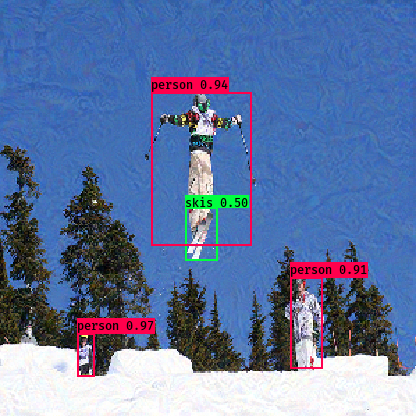}
    \end{minipage}
    \begin{minipage}[b]{0.13\textwidth}
        \includegraphics[width=\textwidth]{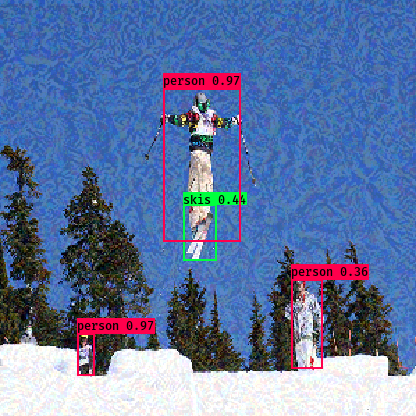}
    \end{minipage}
    \begin{minipage}[b]{0.13\textwidth}
        \includegraphics[width=\textwidth]{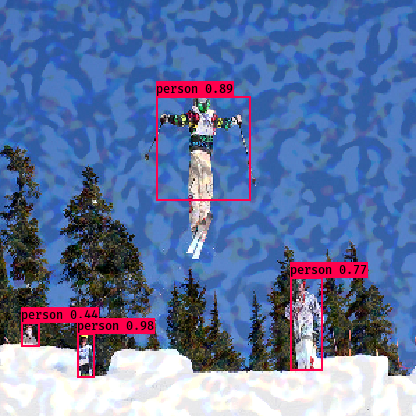}
    \end{minipage}
    \begin{minipage}[b]{0.13\textwidth}
        \includegraphics[width=\textwidth]{det_2_clean}
    \end{minipage}
    \begin{minipage}[b]{0.13\textwidth}
        \includegraphics[width=\textwidth]{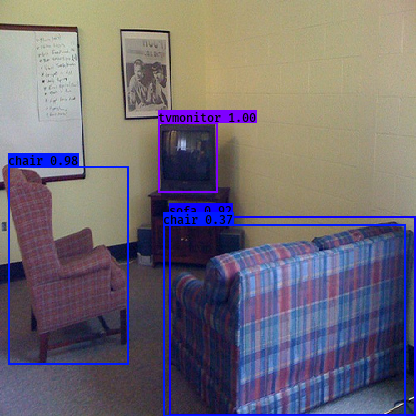}
    \end{minipage}
    \begin{minipage}[b]{0.13\textwidth}
        \includegraphics[width=\textwidth]{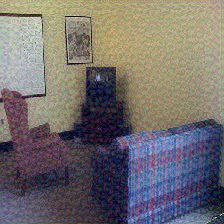}
    \end{minipage}
    \begin{minipage}[b]{0.13\textwidth}
        \includegraphics[width=\textwidth]{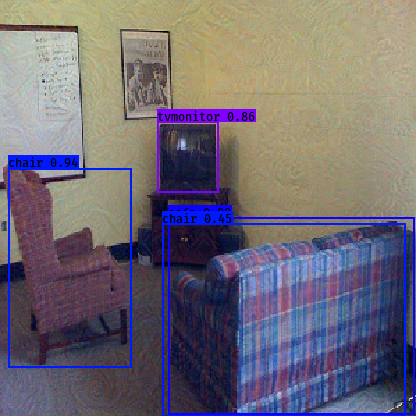}
    \end{minipage}
    \begin{minipage}[b]{0.13\textwidth}
        \includegraphics[width=\textwidth]{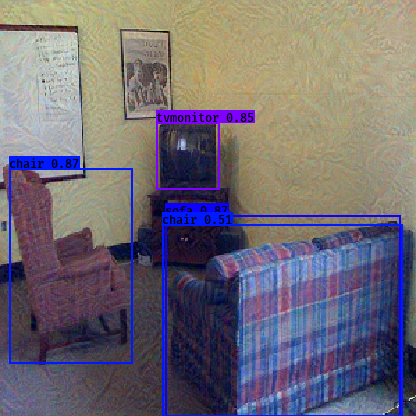}
    \end{minipage}
    \begin{minipage}[b]{0.13\textwidth}
        \includegraphics[width=\textwidth]{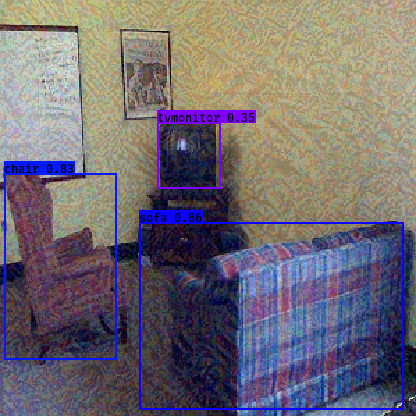}
    \end{minipage}
    \begin{minipage}[b]{0.13\textwidth}
        \includegraphics[width=\textwidth]{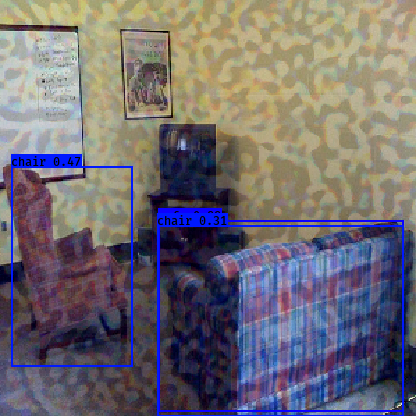}
    \end{minipage}
    \begin{minipage}[b]{0.13\textwidth}
        \includegraphics[width=\textwidth]{seg_1_clean}
    \end{minipage}
    \begin{minipage}[b]{0.13\textwidth}
        \includegraphics[width=\textwidth]{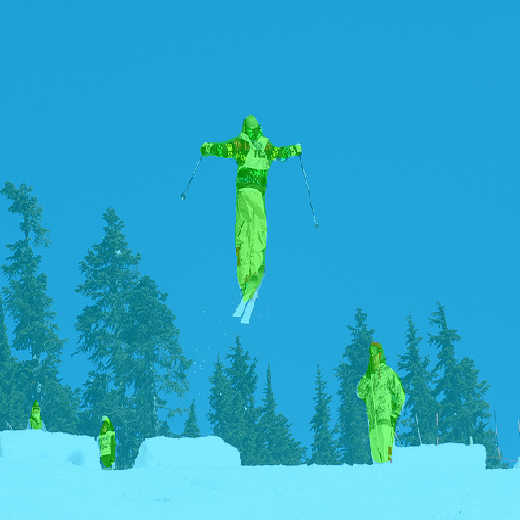}
    \end{minipage}
    \begin{minipage}[b]{0.13\textwidth}
        \includegraphics[width=\textwidth]{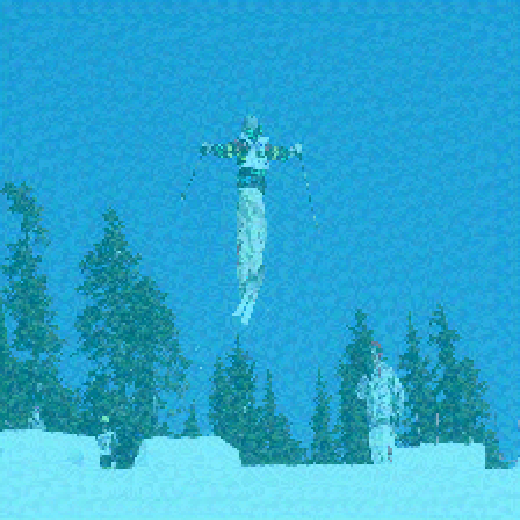}
    \end{minipage}
    \begin{minipage}[b]{0.13\textwidth}
        \includegraphics[width=\textwidth]{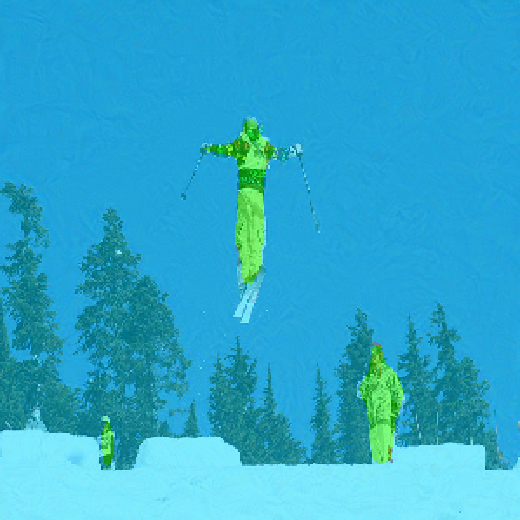}
    \end{minipage}
    \begin{minipage}[b]{0.13\textwidth}
        \includegraphics[width=\textwidth]{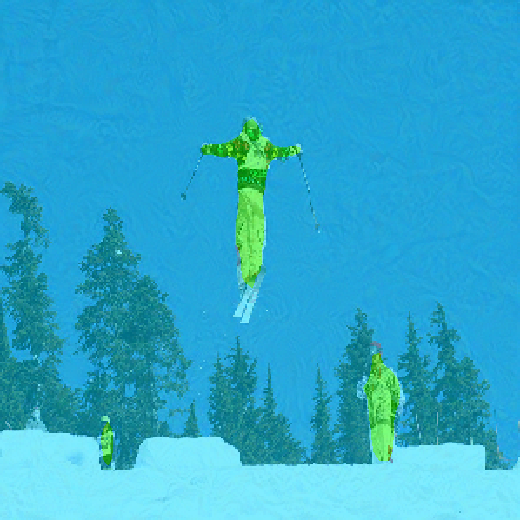}
    \end{minipage}
    \begin{minipage}[b]{0.13\textwidth}
        \includegraphics[width=\textwidth]{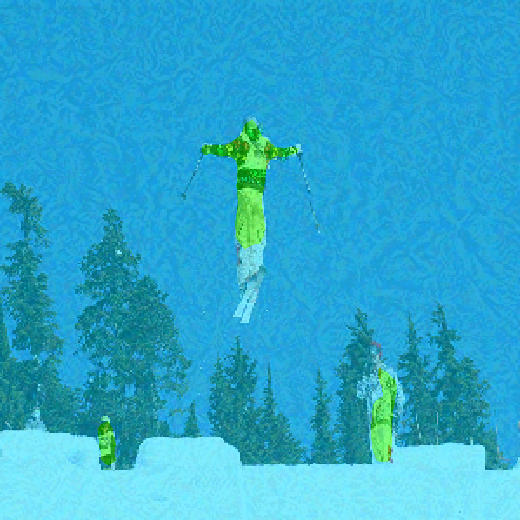}
    \end{minipage}
    \begin{minipage}[b]{0.13\textwidth}
        \includegraphics[width=\textwidth]{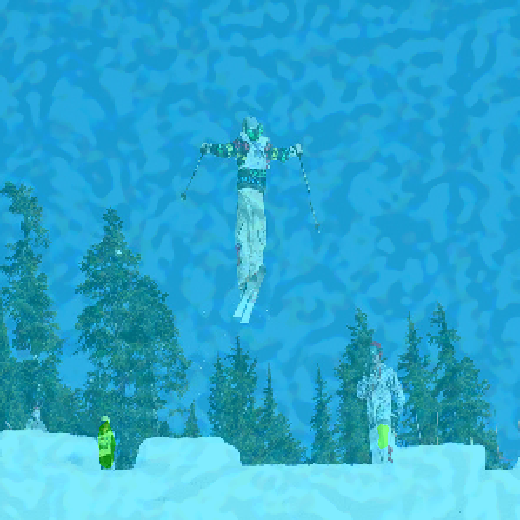}
    \end{minipage}
    \begin{minipage}[b]{0.13\textwidth}
        \includegraphics[width=\textwidth]{seg_2_clean}
        \subcaption{Clean Data}
    \end{minipage}
    \begin{minipage}[b]{0.13\textwidth}
        \includegraphics[width=\textwidth]{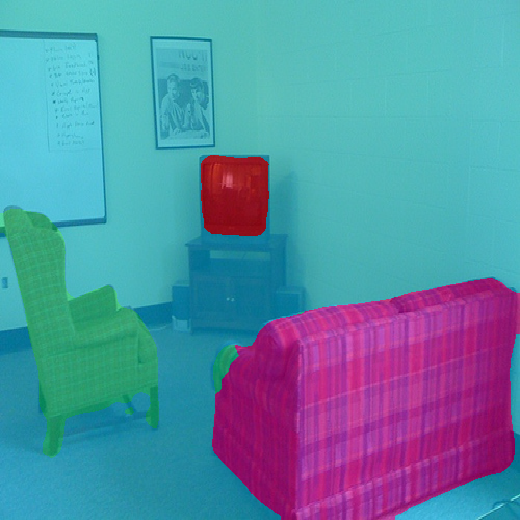}
        \subcaption{Benign}
    \end{minipage}
    \begin{minipage}[b]{0.13\textwidth}
        \includegraphics[width=\textwidth]{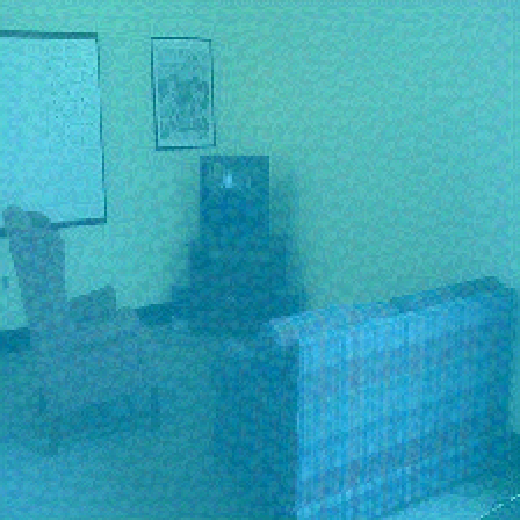}
        \subcaption{DR (ours)}
    \end{minipage}
    \begin{minipage}[b]{0.13\textwidth}
        \includegraphics[width=\textwidth]{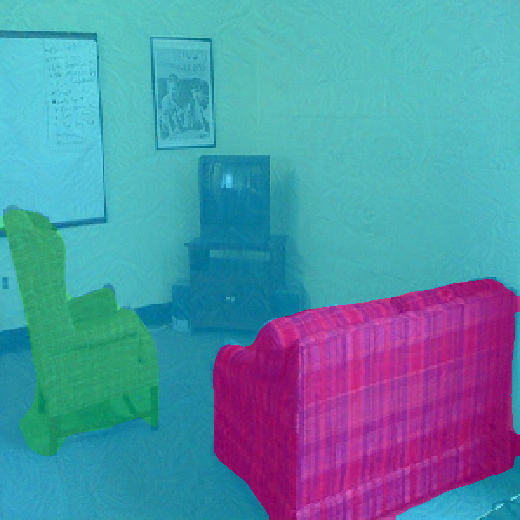}
        \subcaption{PGD}
    \end{minipage}
    \begin{minipage}[b]{0.13\textwidth}
        \includegraphics[width=\textwidth]{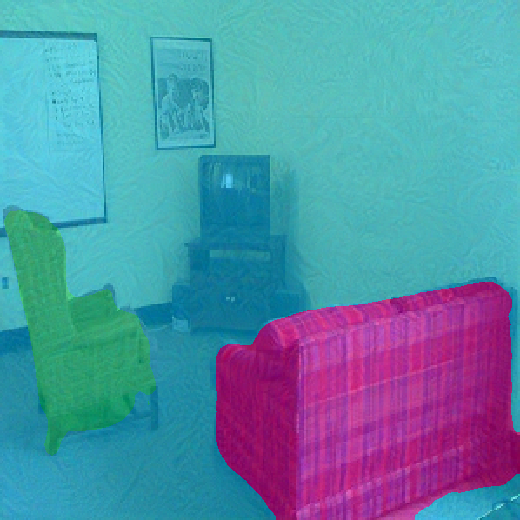}
        \subcaption{MI-FGSM}
    \end{minipage}
    \begin{minipage}[b]{0.13\textwidth}
        \includegraphics[width=\textwidth]{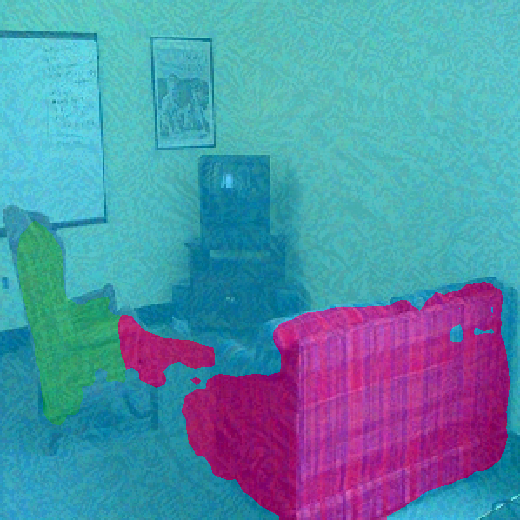}
        \subcaption{DIM}
    \end{minipage}
    \begin{minipage}[b]{0.13\textwidth}
        \includegraphics[width=\textwidth]{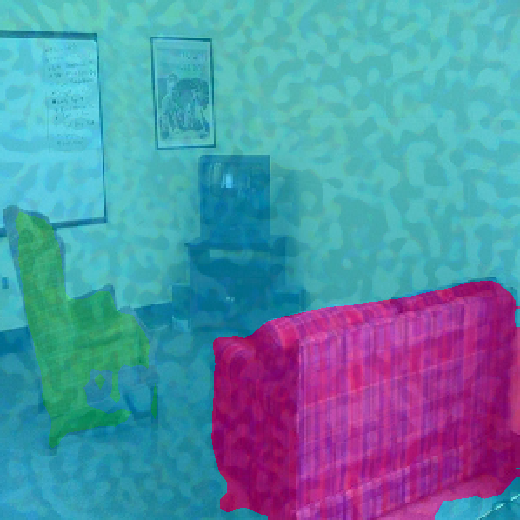}
        \subcaption{TI-DIM}
    \end{minipage}
    \caption{Samples of Detection and Segmentation Results}
    \label{fig:samples}
\end{figure*}

\section{Average Results}

We have compared the proposed DR attack with the state-of-the-art adversarial techniques to demonstrate the transferability of our method on public object detection and semantic segmentation models. We have used the validation sets of ImageNet, VOC2012 and COCO for testing object detection and semantic segmentation tasks. The average results can be seen in Table~\ref{table:avg_det_seg},

For COCO and VOC datasets, our proposed method (\textbf{DR}) achieves the best results by degrading the performance of the target model by a larger margin. For detection, our proposed drops the mAP to 12.8 on average over all the experiments. It creates 3.9 more drop in mAP compared to the best of the baselines (TI-DIM: 16.7 mAP). For semantic segmentation, our proposed attack causes the mIoU to drop to 20.0 on average over all the experiments. It achieves 5.9 more drop in mIoU compared to the best of the baselines (DIM: 25.9 mIoU). 

The diagnostic of average results for ImageNet can be seen in \ref{sec:supp:ImagenetExp}.

\section{Visualization}

Figure~\ref{fig:samples} shows the visualization samples for the proposed method and baselines attacks. Examples of detection and segmentation results for clean images, results for benign images, proposed DR images, PGD images, MI-FGSM images, DIM images and TI-DIM images are shown in each column (starting from left), respectively. First two rows are the detection results, and the last two rows are the segmentation results. We can see that the proposed DR attack is able to effectively perform vanishing attack to both segmentation and detection tasks. It is also noted that the proposed DR attack is more successful and effective, compared to the baselines, when attacking and degrading the performance for smaller objects.

\end{document}